\documentclass[acmsmall,nonacm]{acmart}

\AtBeginDocument{%
  }

\renewcommand\footnotetextcopyrightpermission[1]{}
\makeatletter
\let\@authorsaddresses\@empty
\makeatother

\usepackage{tikz}
\usepackage{braket}
\usepackage{lipsum}
\usepackage{subfig}
\usepackage{xcolor}
\usepackage{amsmath}
\usepackage{balance}
\usepackage{environ}
\usepackage{physics}
\usepackage{wrapfig}
\usepackage{colortbl}
\usepackage{enumitem}
\usepackage{amsfonts}
\usepackage{booktabs}
\usepackage{graphicx}
\usepackage{textcomp}
\usepackage{algorithm}
\usepackage{algpseudocode}
\usepackage[normalem]{ulem}

\newcommand{\sol}{PachinQo}

\newcommand{\rev}[1]{\textcolor{black}{#1}}

\begin{document}

\title{Modeling and Simulating Rydberg Atom Quantum Computers for Hardware-Software Co-design with \sol{}}

\author{Jason Zev Ludmir}\affiliation{\institution{Rice University}\country{Houston, TX, USA}}
\author{Yuqian Huo}\affiliation{\institution{Rice University}\country{Houston, TX, USA}}
\author{Nicholas S. DiBrita}\affiliation{\institution{Rice University}\country{Houston, TX, USA}}
\author{Tirthak Patel}\affiliation{\institution{Rice University}\country{Houston, TX, USA}}

\begin{abstract}

Quantum computing has the potential to accelerate various domains: scientific computation, machine learning, and optimization. Recently, Rydberg atom quantum computing has emerged as a promising quantum computing technology, especially with the demonstration of the zonal addressing architecture. However, this demonstration is only compatible with one type of quantum algorithm, and extending it to compile and execute general quantum algorithms is a challenge. To address it, we propose \sol{}, a framework to co-design the architecture and compilation for zonal addressing systems for any given quantum algorithm. \rev{\sol{}'s evaluation demonstrates its ability to improve a quantum algorithm's estimated probability of success by 45\% on average in error-prone quantum environments.}

\end{abstract}

\maketitle

\section{Introduction}
\label{sec:introduction}

Quantum computing offers the promise of significant speedup compared to transistor-based (classical) computing in areas such as scientific computing, machine learning, and optimization~\cite{preskill2018quantum,preskill2021quantum}. Currently, the hardware for quantum computing is in the development phase, with small-sized computers made available using different technologies that help realize a quantum bit or a qubit~\cite{arute2019quantum,wurtz2023aquila}. Among these technologies, Rydberg atom (also known as a neutral atom) quantum computers, where a qubit is realized using the energy states of a neutral atom, have emerged as a strong candidate~\cite{Sibalic2016-pv,Graham2019-ji,Wintersperger2023,Levine2019-bp,PhysRevLett.104.010502}. Rydberg atom computers offer several advantages over other quantum technologies, primary of which are the ability for qubits to maintain their coherent state for longer runtimes and the ability to move qubits -- as opposed to realizing them on a hardwired chip -- which aids in bringing qubits closer for interactions~\cite{nogrette2014single,Henriet2020-kl,Barredo2018-ie,bluvstein2022quantum}.

Owing to these advantages, a recent demonstration of an error-corrected 280-qubit Rydberg atom quantum computer has shown a promising path toward achieving medium-to-large-sized quantum computers~\cite{bluvstein2024logical}. The demonstration proposed the ``zonal addressing'' architecture for realizing entangling gates that are used to establish quantum interactions among qubits. This architecture divides a Rydberg atom computer into several zones and moves qubits into and out of the entangling (i.e., compute) zone as required. This helps achieve a balance between ``individual addressing''~\cite{shi2020single,patel2022geyser,patel2023graphine,Baker2021-wv,litteken2022reducing}, which is difficult to scale due to the machinery required to address each entangling gate individually, and ``global addressing''~\cite{wurtz2023aquila,bluvstein2022quantum,Wintersperger2023,highfidbluv,tan2022qubit,tan2023compiling}, which is difficult to scale due to its potential for unintended entanglements as it addresses all qubits at all times.

\vspace{2mm}

\noindent\textbf{The Challenge and the Opportunity.} The above proposed zonal addressing architecture was only designed to compile and execute a specific type of quantum code (e.g., surface codes) with all parallel entangling gates~\cite{bluvstein2024logical}. However, in general, quantum computing algorithms have a variety of complex execution patterns, which the architecture needs to accommodate~\cite{li2023qasmbench,patel2020ureqa}. In addition, algorithm executions on quantum computers also suffer from non-negligible hardware errors that are proportional to the number of gates executed and the overall runtime of the quantum ``circuit'' that represents the quantum algorithm~\cite{Henriet2020-kl}. Due to this, the compilation must also be optimized for general quantum algorithms to reduce their overall gate counts and runtimes.

\textit{To address this challenge, we propose \sol{}\footnote{\sol{} is published in the Proceedings of the International ACM SIGMETRICS Conference (SIGMETRICS), 2025.}, a framework to model and simulate the co-design hardware-software of Rydberg atom quantum computers with zonal addressing, optimizing the architecture and compilation for general quantum algorithms.}

\vspace{2mm}

\noindent\textbf{Overview of \sol{}.} \sol{} designs a dual quantum cache that helps store qubits when they do not need to be entangled with other qubits to avoid unintentional entanglement. This dual cache is designed to be compatible with \sol{}'s compiler, which uses it in conjunction with the memory to move qubits in and out of the compute zone rapidly using practical and performant heuristics. \sol{} also carefully initializes the quantum circuit on the quantum computer by leveraging a MaxCut formulation to determine which qubits should be static and which ones should be mobile. Lastly, \sol{} introduces preemptive and parallelizable operations that help prepare qubits for entangling interactions without requiring a runtime increase.

\vspace{2mm}

\noindent\textbf{The contributions of this work are as follows:}

\begin{itemize}[leftmargin=*]
    \item \sol{} is the first-of-its-kind effort to co-design the architecture and compiler for Rydberg atom computers that are compatible with any given quantum algorithm.
    \item \sol{} designs a dual-cache architecture and a heuristics-based compiler, both informed by each other, to create zones for quickly moving qubits for entanglement.
    \item We also develop an easy-to-customize and quick-to-run architectural simulator and compiler implementation, which is open-sourced for the exploration of different architecture and compiler features. \rev{The median algorithm tested compiles within 296.4$ms$ with our implementation.} \rev{\sol{}'s framework is available at: \url{https://github.com/positivetechnologylab/PachinQo}.}
    \item \rev{\sol{}'s evaluation using 13 different quantum algorithms/benchmarks from the QASMBench~\cite{li2023qasmbench} and ArQtiC~\cite{bassman2021arqtic} suites of sizes 51-1000 qubits indicates that \sol{}'s co-design decreases the circuit runtime of an average algorithm by 20\% over competitive techniques.}
    \item \rev{\textit{As a result, \sol{} is also able to improve the estimated probability of success of a quantum algorithm by 45\% on average in error-prone quantum environments.}} Our evaluation also demonstrates that \sol{}'s compiler is adaptable to different static qubit arrangement grids, allowing for optimal algorithm-grid matching without any overhead of tuning the compiler.
\end{itemize}

\section{Brief Relevant Background}
\label{sec:background}

In this section, we provide a brief background on quantum computing principles and Rydberg atom quantum computers.

\vspace{2mm}

\noindent\textbf{Qubits and Quantum States.} Unlike classical bits that only exist in binary states of 0 and 1, quantum computers use qubits that exist in a \textit{superposition} of 0 and 1 simultaneously during computation~\cite{patel2020ureqa}. Each qubit is represented as a vector in Hilbert space, which spans the two-dimensional complex plane. Using bra-ket notation, the state of the qubit is represented as $|\psi\rangle = \alpha|0\rangle + \beta|1\rangle$. $|0\rangle$ and $|1\rangle$ are the two binary states, and $\alpha$ and $\beta$ are two complex coefficients that denote the amplitude and phase corresponding to each state.

With computations involving multiple qubits, the state of an $n$-qubit system is represented as $|\psi\rangle = \sum_{i=0}^{2^n-1} a_i |i\rangle$. For example, a two-qubit system exists in a superposition of the following four states: $|00\rangle$, $|01\rangle$, $|10\rangle$, and $|11\rangle$. Upon \textit{measurement}, the state of a quantum system collapses and manifests as one of the $2^n$ states. The probability of measuring state $i$ in an $n$-qubit system is $|a_i|^2$. Thus, $\sum_{i=0}^{2^n-1} |a_i|^2 = 1$. As a result, the output of a quantum algorithm is a probability distribution.

\begin{wrapfigure}{r}{0.65\textwidth}
    \vspace{-7mm}
    \centering
    \subfloat[Example Quantum Circuit]{\includegraphics[scale=0.54]{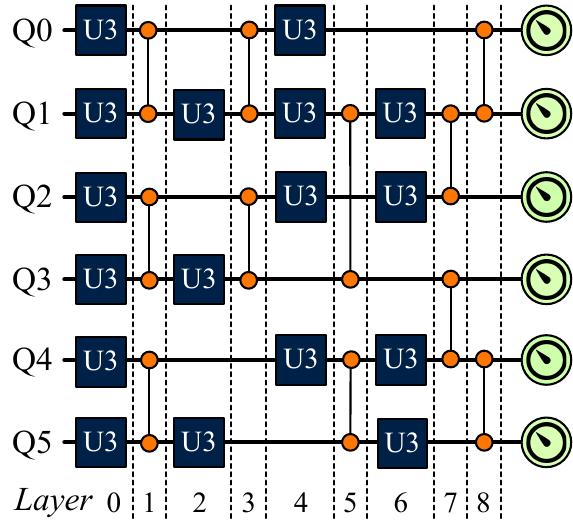}}
    \hfill
    \subfloat[Rydberg Atom Tech.]{\includegraphics[scale=0.38]{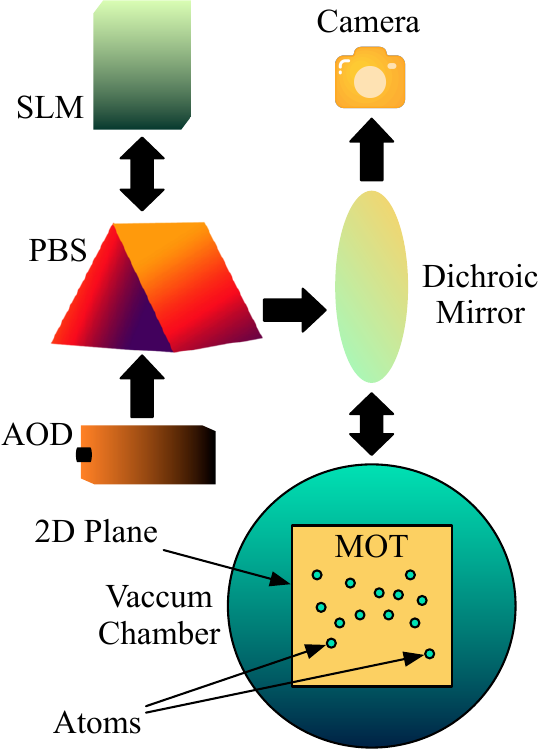}}
    \vspace{1mm}
    \hrule
    \vspace{-3.5mm}
    \caption{(a) An example quantum circuit. The qubit state evolves from left to right (horizontal lines); the vertical lines with circular ends represent the CZ gates between the two corresponding qubits. At the end of the circuit, all qubits are measured using readout operations (circular meters). (b) Physical setup of a Rydberg atom quantum computer.}
    \vspace{-2mm}
    \label{fig:circuit}
\end{wrapfigure}

\vspace{2mm}

\noindent\textbf{Quantum Gates and Circuits.} Quantum states are manipulated using quantum gates, which are represented mathematically by unitary matrices that act on the state vector. In general, a gate acting on an $n$-qubit state corresponds to a $2^n\times2^n$ matrix. One particular example is the U3 gate, a one-qubit gate that puts a qubit in a desired superposition. To make qubits interact and \textit{entangle} them, we can use the two-qubit controlled-Z (CZ) gate.

Together, U3 and CZ form a universal basis set, meaning we can reduce any given quantum gate into these two gates~\cite{patel2022geyser}. A sequence of gates applied to a quantum state is referred to as a quantum algorithm or circuit, an example of which is shown in Fig.~\ref{fig:circuit}(a). \rev{One additional gate relevant to our discussion is the SWAP gate, which swaps the states of two qubits; the gate can be decomposed into three CZ gates and six U3 gates}. It does not alter the logic of a circuit, but it needs to be inserted in a circuit in specific scenarios, as we discuss later in this section.

When a quantum algorithm is executed once, it generates one output state according to the probability associated with it. Thus, a quantum algorithm has to be run multiple times to generate a full output probability distribution over each state, which is then interpreted as per the application of the algorithm. All quantum computers are attached to a classical computer, which interfaces with the quantum computer, provides the control logic, schedules the circuit gates, receives the output states, and interprets the output distribution.

\begin{wrapfigure}{r}{0.65\textwidth}
    \vspace{-5mm}
    \centering
    \subfloat[Interaction/Blockade Radius]{\includegraphics[scale=0.3]{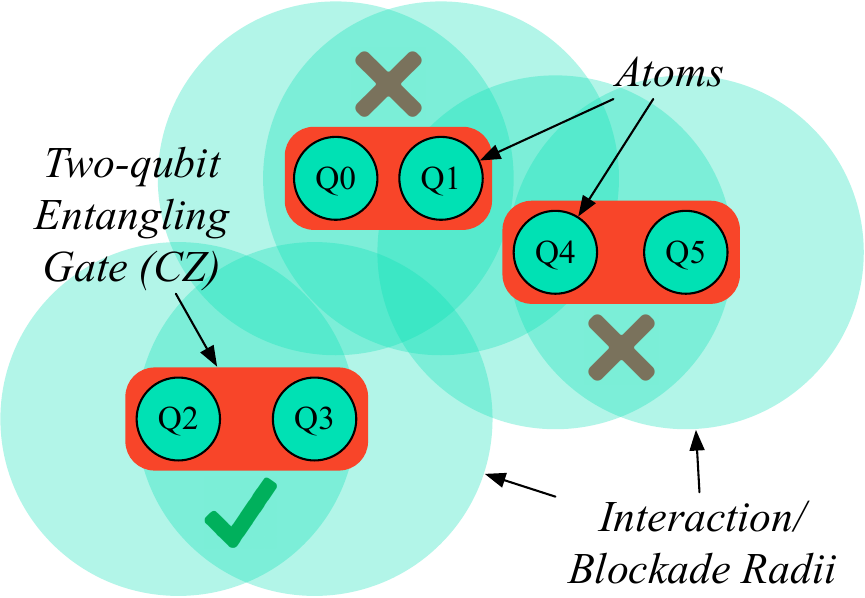}}
    \hfill
    \subfloat[SLM vs. AOD Atoms]{\includegraphics[scale=0.3]{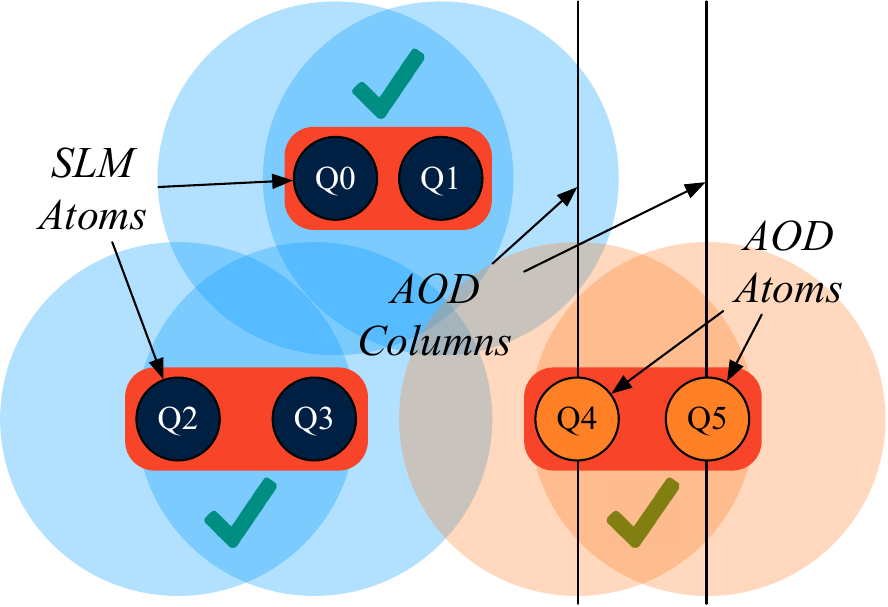}}
    \vspace{1mm}
    \hrule
    \vspace{-3.5mm}
    \caption{(a) The CZ gate between qubits Q2-Q3 can execute as they are within each other's interaction radius and are not blocked by other qubits. However, the CZ gates between Q0-Q1 and Q4-Q5 cannot execute as Q1 and Q4 are blocked by each other. (b) By putting Q4 and Q5 in the AOD (the rest can be in the SLM), they can be moved farther away from Q0-Q1, and now all three CZ gates can execute in parallel.}
    \vspace{-1mm}
    \label{fig:constraints}
\end{wrapfigure}

\vspace{2mm}

\noindent\textbf{Rydberg Atom Quantum Computers.} In a Rydberg atom computer, qubits are formed from the energy levels of valence electrons belonging to neutral atoms like Cesium or Rubidium~\cite{Sibalic2016-pv,Graham2019-ji}. The atoms used have one valence electron, so each atom provides one qubit. Therefore, in this paper, we use ``atom'' and ``qubit'' interchangeably. Two levels near the ground state are chosen as the $\ket{0}$ and $\ket{1}$ states. In addition to these two states, an additional high energy state with a large quantum number is used during certain logical operations~\cite{wurtz2023aquila}. This state is referred to as the Rydberg state.

As shown in Fig.~\ref{fig:circuit}(b), atoms are cooled in a magneto-optical trap (MOT), then trapped using a spatial light modulator (SLM) or an acousto-optic deflector (AOD) via the polarizing beam splitter (PBS)~\cite{Levine2018-ma,Levine2019-bp,Henriet2020-kl,saffman2020symmetric}. \textit{The SLM creates an array of stationary optical tweezer traps in a specified configuration, while the AOD provides a movable array of optical traps.} Operations on the qubit-atoms are conducted with the use of laser pulses, and post-computation, a fluorescence-sensing camera reads out the atom array to measure the qubit state~\cite{Adams2019-sc,Henriet2020-kl}.

\vspace{2mm}

\noindent\textbf{Rydberg Atom Challenges.} One-qubit U3 gates are applied using Raman pulses. To apply the two-qubit CZ gate, there needs to be an interaction between the two qubits. Rydberg atom systems use the dipole-dipole interaction between atoms in the Rydberg state. This interaction's strength scales inversely to the separation between atoms. This coupling between two Rydberg atoms in \textit{interaction radius} can be modeled as a CZ gate~\cite{Henriet2020-kl}, as shown in Fig.~\ref{fig:constraints}(a) for the Q2-Q3 CZ gate. However, this also means that other atoms cannot exist within the interaction radius, as they may get unintentionally entangled. Thus, the interaction radius also doubles as a \textit{blockade radius}. In Fig.~\ref{fig:constraints}(a), Q1 and Q4 block each other, and therefore gates Q0-Q1 and Q4-Q5 cannot execute.

In cases like this, as shown in Fig.~\ref{fig:constraints}(b), qubits Q4 and Q5 can be placed in the AOD traps and moved farther away so that all three pairs are outside of each other's interaction/blockade radii. However, there are several constraints associated with AOD movement~\cite{bluvstein2024logical}: (1) AOD traps exist as columns/rows as they are laser-based traps. Thus, all atoms placed within an AOD column/row must move in tandem with each other as the whole column/row moves. (2) AOD traps cannot overlap and move past each other as the laser traps cannot interfere with each other. Thus, the relative ordering of AOD columns/rows is always maintained. 

\vspace{2mm}

\begin{figure*}[t]
    \centering
    \includegraphics[scale=0.26]{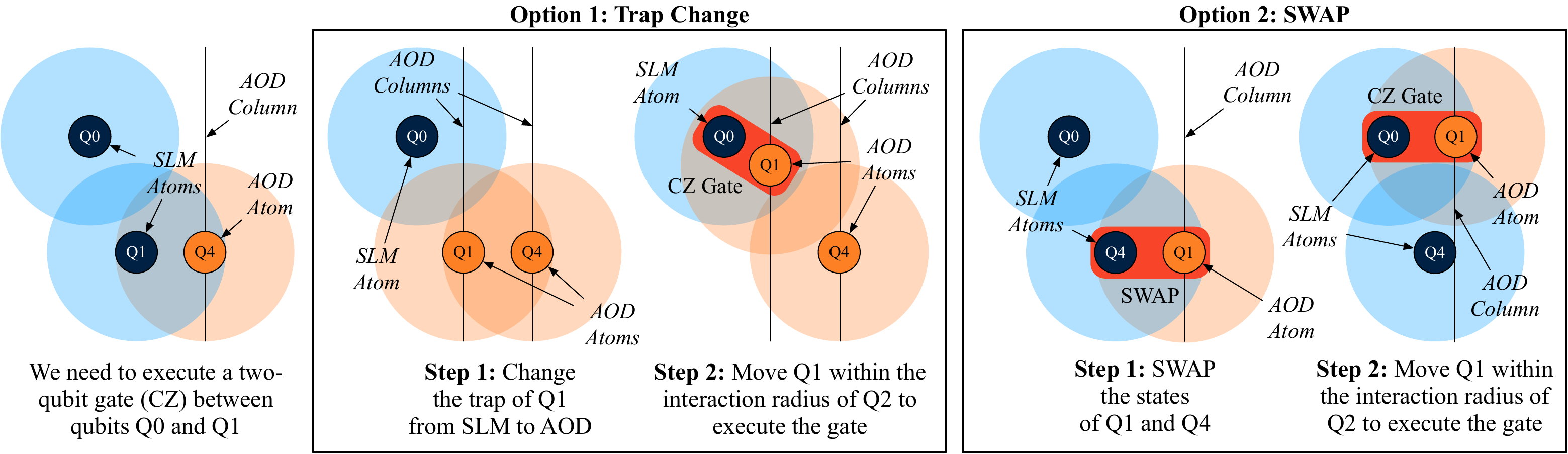}
    \vspace{1mm}
    \hrule
    \vspace{-3.5mm}
    \caption{Two methods of executing the Q0-Q1 CZ gate: (1) Trap change to convert Q1 from an SLM to an AOD qubit, and (2) SWAP the states of Q1 and Q4. Trap changes are time-consuming, and SWAPs are error-prone.}
    \vspace{-1mm}
    \label{fig:trap_swap}
\end{figure*}

\noindent\textbf{Trap Changes and SWAPs.} Switching an atom from one type of trap to another (e.g., SLM to AOD or AOD to SLM) is referred to as a trap change. Trap changes are used in specific initialization and mid-computation scenarios. Consider the example shown in Fig.~\ref{fig:trap_swap}, where SLM qubit Q1 needs to be moved closer to SLM qubit Q0 to run a CZ gate between them. In this case, one option would be to perform a trap change operation on Q1 to convert it into an AOD qubit and then move it closer to Q0 for the CZ gate. The second option would be to run a SWAP operation between Q1 and Q4 (which is an AOD qubit) to swap their states and then move Q1 (which is now in the AOD) close to Q0 for the CZ gate. Both options are sub-optimal: (1) A trap change is time-consuming (takes 25$\times$ the amount of time as a CZ gate), but (2) \rev{A SWAP is error-prone (has $>3\times$ the error of a CZ gate). Thus, \sol{} must minimize both in its design.}

\vspace{1mm}

That concludes our discussion of background. Next, we discuss the motivation for \sol{}.



\begin{figure*}[t]
    \centering
    \includegraphics[scale=0.45]{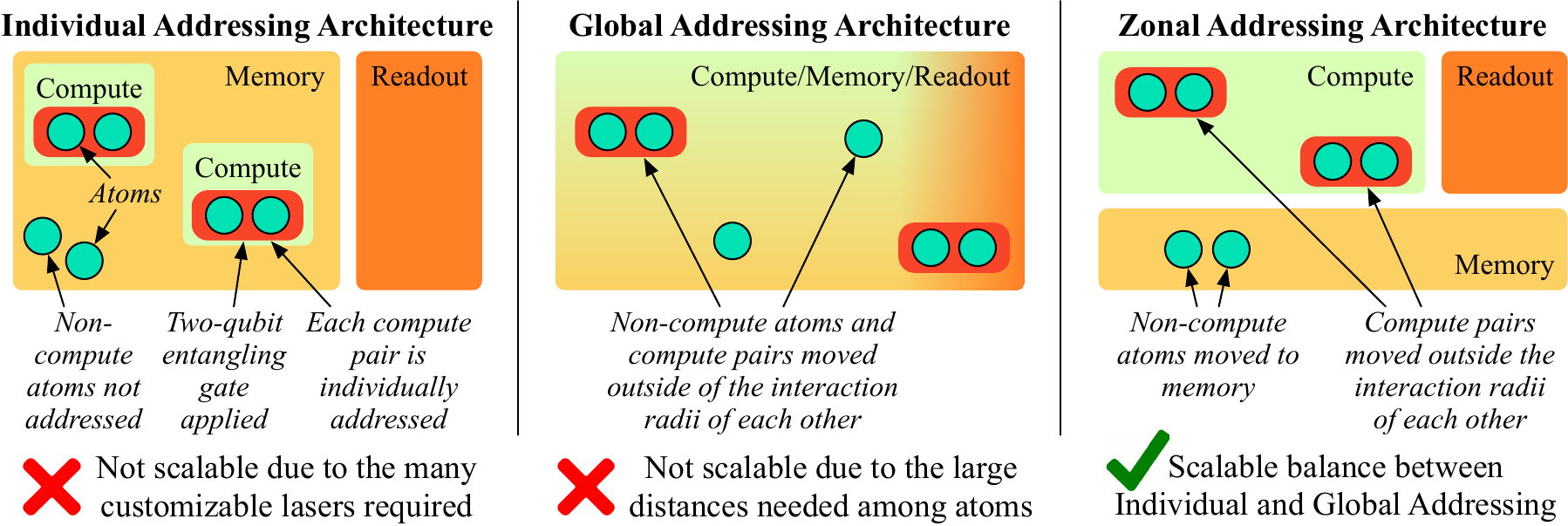}
    \vspace{0.5mm}
    \hrule
    \vspace{-3.5mm}
    \caption{We develop \sol{} for the recently-demonstrated zonal addressing architecture~\cite{bluvstein2024logical}, which achieves the best balance between the individual and global addressing architectures, making it a promising scalable configuration. \textbf{\textit{The zones are not depicted to scale -- their dimensions vary in different figures based on ease of visualization. In our experiments, we maintain consistent and realistic scales.}}}
    \vspace{-1mm}
    \label{fig:addressing}
\end{figure*}
\section{Motivation for \sol{}}
\label{sec:motivation}

In most prior work, compilation and architecture exploration techniques were designed for Rydberg atoms systems with either \textit{individual} Rydberg addressing~\cite{shi2020single,patel2022geyser,patel2023graphine,Baker2021-wv,litteken2022reducing} or \textit{global} Rydberg addressing~\cite{wurtz2023aquila,bluvstein2022quantum,Wintersperger2023,highfidbluv,tan2022qubit,tan2023compiling}. As Fig.~\ref{fig:addressing} shows, an individual addressing system would be ideal since, with such a computer, users could arbitrarily select qubit pairs that would get illuminated by the Rydberg laser at arbitrary time intervals -- referred to as a \textit{compute} unit, while the remaining qubits are in \textit{memory} (qubits are moved to \textit{readout} for measurement). However, this is very difficult to implement in practice due to the fact that simultaneously addressing different qubits locally would involve many Rydberg lasers in the same computer (in fact, as many Rydberg lasers as half the number of qubits in the system since each pair of qubits could execute a CZ gate in parallel). 

With global addressing, the entire grid of atoms is illuminated simultaneously, meaning that if any atoms are within the Rydberg interaction radius of each other, they will entangle. This is a severe limitation as it makes it very difficult to position atoms such that they do not interact unintentionally (known as crosstalk). The implication of this is that with global Rydberg addressing, it is difficult to compile quantum programs since every time any entanglement is desired, all qubits must be accounted for and moved far away from all other qubits that are not supposed to entangle. 

As a median between individual and global addressing, recently, the idea of \textit{zonal} addressing has been demonstrated for Rydberg atom systems for a 280-qubit Rydberg atom system~\cite{bluvstein2024logical}. The idea with zonal addressing is that, like global addressing, an entire compute area is illuminated by the Rydberg laser. Unlike global addressing, this compute zone only represents a subset of the total grid space in the Rydberg atom system. This means that atoms that are not in the compute zone do not cause or suffer crosstalk from Rydberg excitation. The advantages of the zonal architecture are twofold: it is more feasible to implement than an individual addressing system, and it offers greater flexibility compared to a global system. This is because non-compute atoms can be moved out of the compute area and into the memory area where no Rydberg laser is applied.

\begin{figure}[t]
    \centering
    \includegraphics[scale=0.4205]{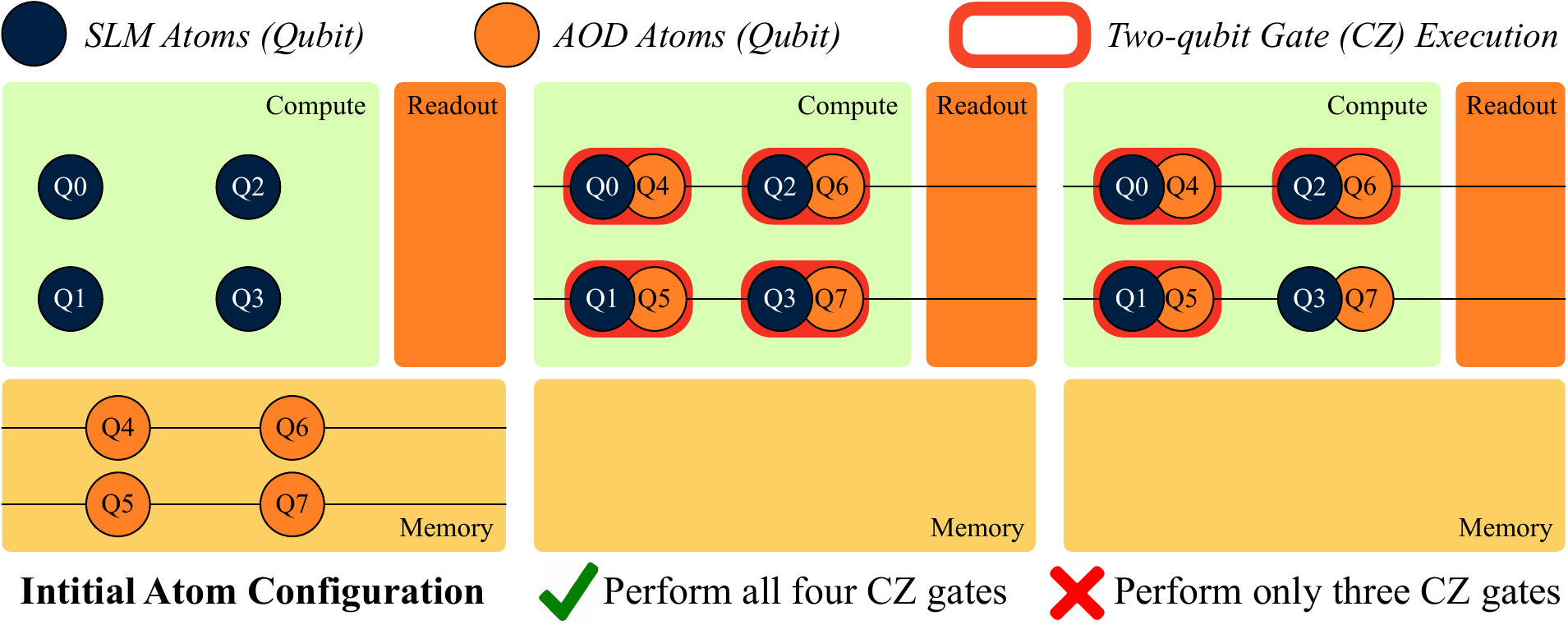}
    \vspace{0.5mm}
    \hrule
    \vspace{-3.5mm}
    \caption{In the current zonal architecture, it is possible to run algorithms with all parallel two-qubit CZ gates (e.g., surface codes) but not general quantum computing algorithms.}
    \vspace{-1mm}
    \label{fig:motive}
\end{figure}

However, research on compiling quantum programs for Rydberg atom systems has largely focused on individual and global addressing~\cite{wurtz2023aquila,bluvstein2022quantum,PhysRevA.105.032618,Wintersperger2023,highfidbluv,Wintersperger2023,shi2020single}. This is because zonal addressing, a newer architecture, has only recently been introduced~\cite{bluvstein2024logical}. Moreover, the only research on compilation for zonal addressing architecture done thus far does not introduce any generalized compilation/architecture strategies; rather, it focuses on implementing a specific algorithm for quantum error correction or surface codes~\cite{bluvstein2024logical,stade2024abstractmodelefficientrouting}. Surface codes only have parallel quantum CZ gates in distinct patterns, as shown in the example snapshot shown in Fig.~\ref{fig:motive}. Consider the four SLM qubits that need to be entangled with the four AOD qubits. In the shown example, the four qubits can be put on two AOD rows and brought close to the SLM qubits to execute the CZ gates.

On the other hand, general quantum algorithms do not have all parallel gates. For example, if only three of the CZ gates need to be run (Q0-Q4, Q2-Q6, and Q1-Q5), that is not possible with this compilation, as Q7 has to be moved in tandem with Q5. Thus, general quantum algorithms require more sophisticated compilation techniques that can cater to algorithm-specific execution scenarios. Accordingly, the architecture should also be optimized for general quantum algorithms. \textit{To achieve this, our framework, \sol{}, co-designs the architecture and compilation of general quantum algorithms for zonal addressing Rydberg atom computers.}

\section{Design and Implementation}
\label{sec:design}

In this section, we provide a detailed description of \sol{}'s design and implementation. We begin by proposing a change to the zonal addressing architecture.

\subsection{The Quantum Cache}

Zonal addressing computers typically have three types of zones with differing purposes: the memory zone, which is used to store atoms in bulk so that they do not entangle; the compute zone, where the Rydberg laser illuminates the whole zone to entangle atoms within close proximity of each other; and the readout zone, where the fluorescence-sensing camera measures the final states of atoms. The main goal when utilizing the different zones is maximizing the parallel execution of gates when running a given quantum circuit while minimizing the effects of crosstalk in compute. This goal is complicated by the fact that AOD movement is heavily restricted by two non-trivial constraints: (1) atoms in each column move in tandem, and (2) the relative order of AOD columns along the X-axis must be maintained throughout the execution of the circuit -- AOD columns cannot overlap and move past each other.

\begin{figure}[t]
    \centering
    \includegraphics[scale=0.6]{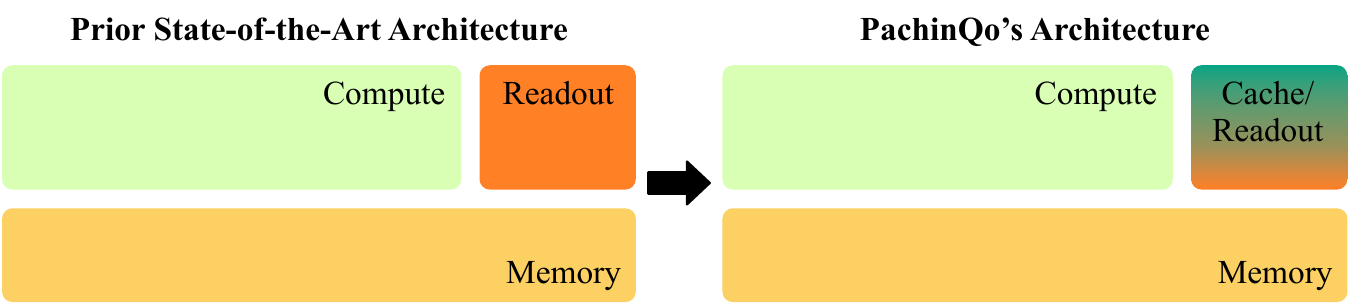}
    \vspace{1mm}
    \hrule
    \vspace{-3.5mm}
    \caption{We propose the use of the readout zone as a cache to store AOD qubits out of the compute zone.}
    \vspace{-1mm}
    \label{fig:cache}
\end{figure}

In order to minimize crosstalk during the execution of a quantum circuit, we want to have locations to store unused AOD columns outside of compute. Just using a single memory zone for compilation is non-optimal since it is very difficult to maneuver AOD columns into desired locations in compute when some of the columns might not be used while others might be (i.e., general quantum algorithms with non-parallel CZ gates), and the unused columns block movement along the X-axis due to the relative ordering constraint. This is where \sol{} introduces the idea of a \textbf{quantum cache}. The quantum cache is a zone where AOD columns not being used for CZ gate execution can be temporarily stored. As shown in Fig.~\ref{fig:cache}, \sol{} uses the readout zone for this, which can be doubled as a cache when the qubits are not being measured without requiring additional area. In prior work, the readout zone has been used solely for measurements at the circuit end~\cite{bluvstein2024logical}. However, using the zone solely for measurement means it is idle during circuit execution and thus can be used as a cache. Having multiple zones to store non-compute qubits allows for adaptable temporary atom movement that minimizes the effects of the relative ordering constraint. The benefits of having a quantum cache will become clearer as the design of \sol{}'s compilation is fleshed out, starting with the initialization of the zonal architecture for a given quantum algorithm.

\begin{figure*}[t]
    \centering
    \includegraphics[scale=0.2905]{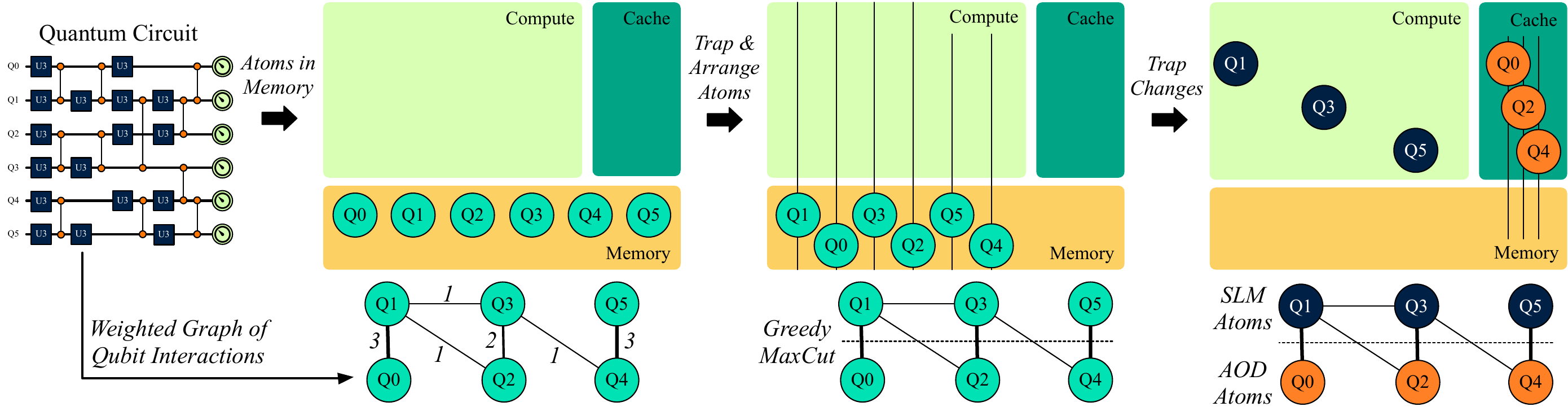}
    \vspace{0.5mm}
    \hrule
    \vspace{-3.5mm}
    \caption{Visual representation of \sol{}'s procedure for mapping the circuit onto the qubits and initializing the qubit types and positions in the zonal architecture.}
    \vspace{-1mm}
    \label{fig:init}
\end{figure*}

\subsection{System Initialization for a Quantum Algorithm}

For a given quantum algorithm, \sol{} begins by loading the appropriate number of qubits based on the algorithm's circuit into the memory zone, where all atoms are trapped in static SLM traps (see ``Atoms in Memory'' step in Fig.~\ref{fig:init}). Note that this is a hardware constraint, as all atoms must first be loaded into the memory. Qubits in the quantum circuit are arbitrarily mapped to atoms in memory since, at this point, all atoms are functionally identical.

Recall that one of the goals of \sol{} is to maximize parallelism in compute, namely by executing CZ gates in parallel to reduce the overall circuit runtime. To execute a CZ gate, two atoms must be close together (e.g., within interaction radius), but given the constraints on AOD column movement, it can be very challenging to get any two columns in the AOD close together. If there were any other columns between the two columns of interest (e.g., if the columns were not adjacent), these ``sandwiched'' columns would block the execution of the CZ gate. Thus, we would have to execute a series of SWAP gates to get the qubits into atoms in adjacent AOD columns. \rev{Not only is performing SWAPs time-consuming but recall that SWAP gates each consist of three high-error sequential CZ operations, each with roughly 0.5\% error (along with six low-error U3 gates), which would significantly increase the error of our quantum circuit.} Thus, another goal of \sol{} is to minimize the number of SWAPs.

One strategy to minimize SWAPs involves placing some atoms in static SLM traps within the compute zone. Since SLM traps do not hinder the movement of AOD columns, employing these static traps simplifies the task of arranging AOD columns without the challenges of relative column ordering. However, relying solely on SLM traps is impractical as none of the qubits would be able to move during execution. The entire compute zone is illuminated by the Rydberg laser, causing all SLM-trapped atoms that are within the interaction radius of each other to entangle whenever the laser is activated. To avoid undesired crosstalk, these SLM atoms must be distanced from one another. Therefore, \sol{} cannot depend entirely on either mobile AOD columns, due to their movement constraints, or on static SLM traps, which would limit the execution of quantum circuits. In fact, by ensuring that for any CZ gate, one qubit would be in an SLM trap and another in an AOD column, \textit{we decrease the need for extensive SWAP operations while at the same time reducing the impact of AOD columns interfering with one another}.

\begin{algorithm}[t]
\caption{\rev{Greedy MaxCut Grouping Algorithm}}
{\small
\begin{algorithmic}[1]
\State $QC \gets $ Quantum Circuit, $G_{AOD} \gets $ $\{\phi\}$, $G_{SLM} \gets $ $\{\phi\}$
\State \textbf{for} each CZ gate $g \in QC$ 
\State \ \ \textbf{for} qubits $q_1,q_2 \in g$
\State \ \ \ \ Find qubit to atom mapping $q_1,q_2 \rightarrow a_1,a_2$
\State \ \ \ \ \textbf{if} $a_1 \notin (G_{AOD} \cup G_{SLM})$ and $a_2 \notin (G_{AOD} \cup G_{SLM})$
\State \ \ \ \ \ \ $G_{AOD}.\text{append}(a_1)$
\State \ \ \ \ \ \ $G_{SLM}.\text{append}(a_2)$
\State \ \ \ \ \textbf{else if} $a_1 \in G_{\text{AOD}}$ \textbf{then} $G_{\text{SLM}}.\text{append}(a_2)$
\State \ \ \ \ \textbf{else if} $a_1 \in G_{\text{SLM}}$ \textbf{then} $G_{\text{AOD}}.\text{append}(a_2)$
\State \ \ \ \ \textbf{else if} $a_2 \in G_{\text{AOD}}$ \textbf{then} $G_{\text{SLM}}.\text{append}(a_1)$
\State \ \ \ \ \textbf{else if} $a_2 \in G_{\text{SLM}}$ \textbf{then} $G_{\text{AOD}}.\text{append}(a_1)$
\State \ \ \textbf{return} $G_{AOD}$, $G_{SLM}$
\end{algorithmic}}
\label{algo:maxcut}
\end{algorithm}

Thus, \sol{} splits qubits into two groups: (1) one that is mobile in the AOD columns and (2) one that is static in SLM traps within compute. The goal is to group the atoms such that \sol{} maximizes the number of CZ gates with one qubit in one group and one qubit in the other group, which we formulate as the \textit{maximum cut or MaxCut} problem~\cite{commander2009maximum}. The MaxCut problem involves cutting a graph into two partitions such that the sum of the weights of the edges crossing the cut is maximized. This ensures that most gates have one qubit in the SLM and the other in the AOD (see Fig.~\ref{fig:init}). As a practical, greedy heuristic for this NP-complete problem, \sol{} begins by analyzing each two-qubit CZ gate in the input quantum circuit (Algorithm~\ref{algo:maxcut}). For each CZ gate, \sol{} checks to see if either qubit has thus far been placed in the AOD or the SLM. If neither qubit has yet been placed in either group, then one qubit is placed in the SLM and one in the AOD. If one of the CZ qubits has already been grouped but the other has not, then the non-grouped qubit is placed in the complementary group (e.g., if the already-grouped qubit is in the SLM, the other qubit is placed in the AOD group, and vice versa). If a CZ gate is identified where both qubits have been grouped, there is nothing to be done. If we try to place a qubit in one group but there is no more room in the group (meaning there is insufficient space in the AOD or the SLM to avoid crosstalk concerns), we group the qubit into whichever group still has room. By the end of this procedure, all qubits are grouped into the AOD or the SLM.

Once the atoms have been grouped, the SLM group atoms (which are currently trapped statically in the memory's SLM) are transferred to the AOD, moved into compute, and transferred again from the AOD into the static SLM compute traps. Recall that since switching atom traps between the AOD and SLM devices is a time-intensive operation, we want to minimize the number of sequential trap changes. Since a trap change takes about 156 times as long as the processing of a CZ gate, performing these trap changes sequentially could take considerable time. For example, if 64 atoms need to be transferred from memory to compute, doing 64 sequential trap changes would take about as long as performing 1600 sequential CZ gates, making the process highly inefficient. \sol{} implicitly guarantees that all the trap changes required to transfer all necessary atoms from the memory SLM into the AOD can be run in parallel, and all the trap changes required to transfer all atoms from the AOD into the computer's SLM can also be run in parallel. This guarantee exists from the fact that \sol{} initially organizes the atoms in memory into columns; since the AOD is similarly structured into columns, it simply is a matter for \sol{} to position the AOD columns adjacent to the memory columns, and switch all of the desired atoms into the AOD. \sol{}'s parallelized trap change is thus an important optimization as it ensures that only two serial trap changes are required during the initialization of the SLM atoms regardless of the number of qubits in the quantum algorithm being executed.

Once the SLM atoms are in place, then the remainder of the atoms in memory are placed in the AOD -- parallel trap changes, equivalent to one serial trap change, are required here. The AOD columns, now loaded with atoms, are moved into the cache zone (see the final stage of Fig. ~\ref{fig:init}). At this point, the initialization process is complete, and \sol{} is ready to begin the circuit execution. However, to optimize the execution, we propose another architectural change.

\begin{figure}[t]
    \centering
    \includegraphics[scale=0.6]{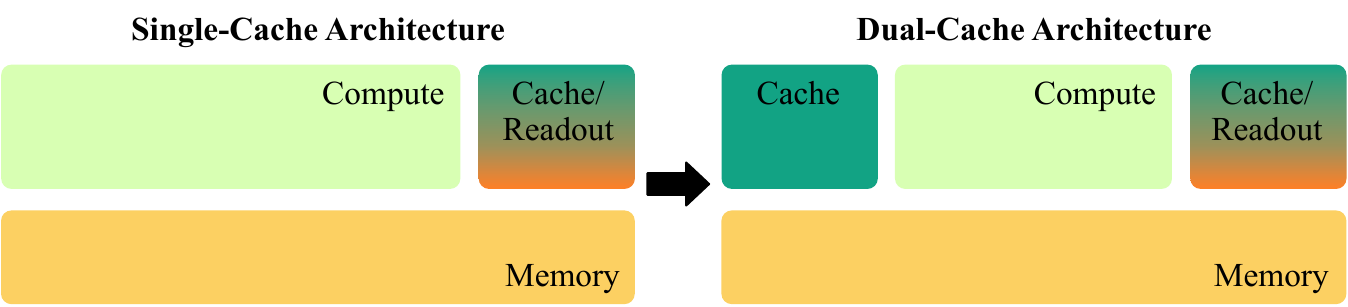}
    \vspace{1mm}
    \hrule
    \vspace{-3.5mm}
    \caption{We propose an architecture with two caches, one on both sides of the compute zone, for ease of AOD movement. The right cache still doubles as the readout zone.}
    \vspace{-1mm}
    \label{fig:two_caches}
\end{figure}

\begin{figure*}[t]
    \centering
    \includegraphics[scale=0.3105]{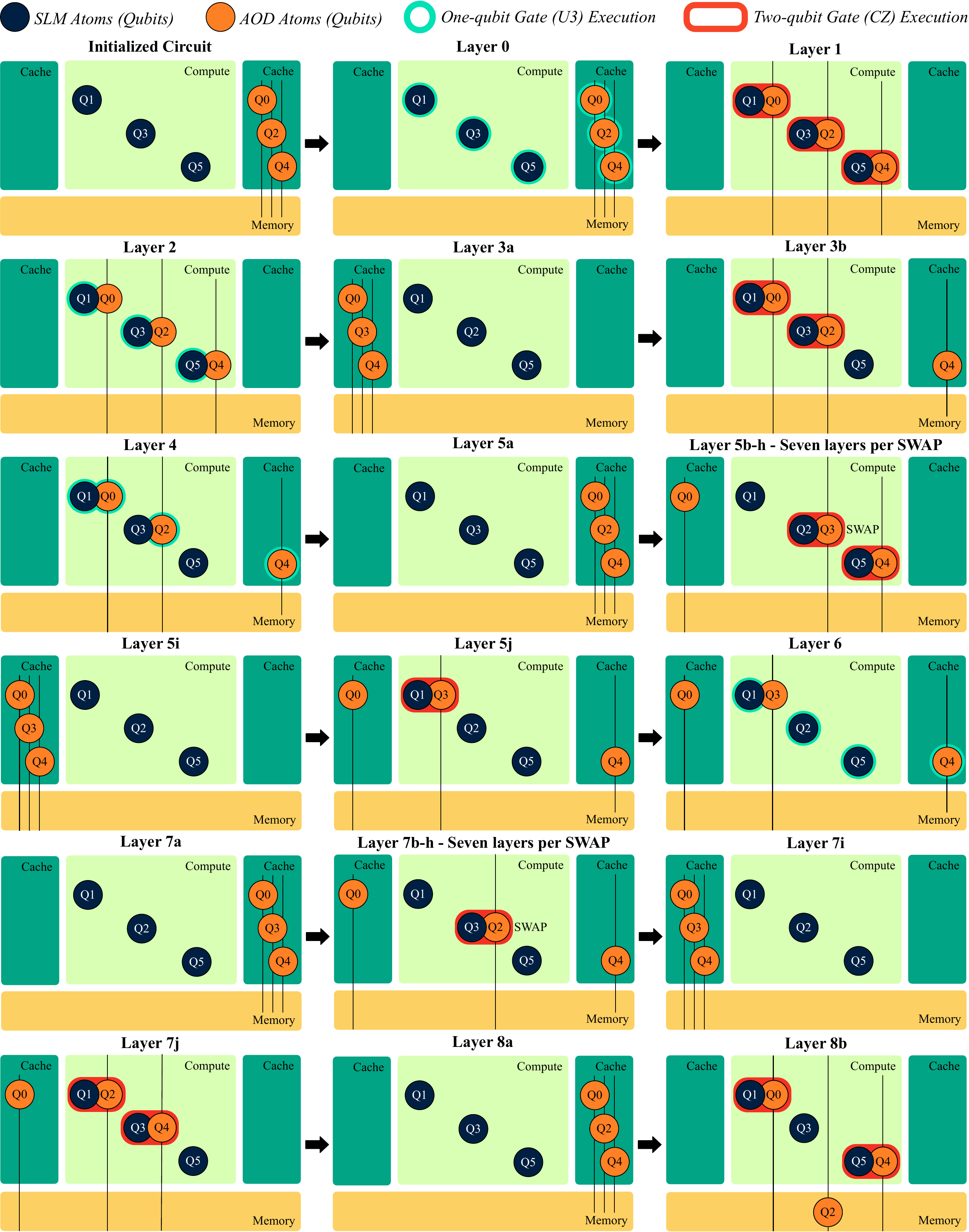}
    \vspace{1mm}
    \hrule
    \vspace{-3.5mm}
    \caption{\rev{The steps involved in \sol{}'s layer-by-layer execution of the example circuit shown in Fig.~\ref{fig:circuit}(a).}}
    \vspace{-1mm}
    \label{fig:execution}
\end{figure*}

\begin{figure}[t]
    \centering
    \includegraphics[scale=0.44]{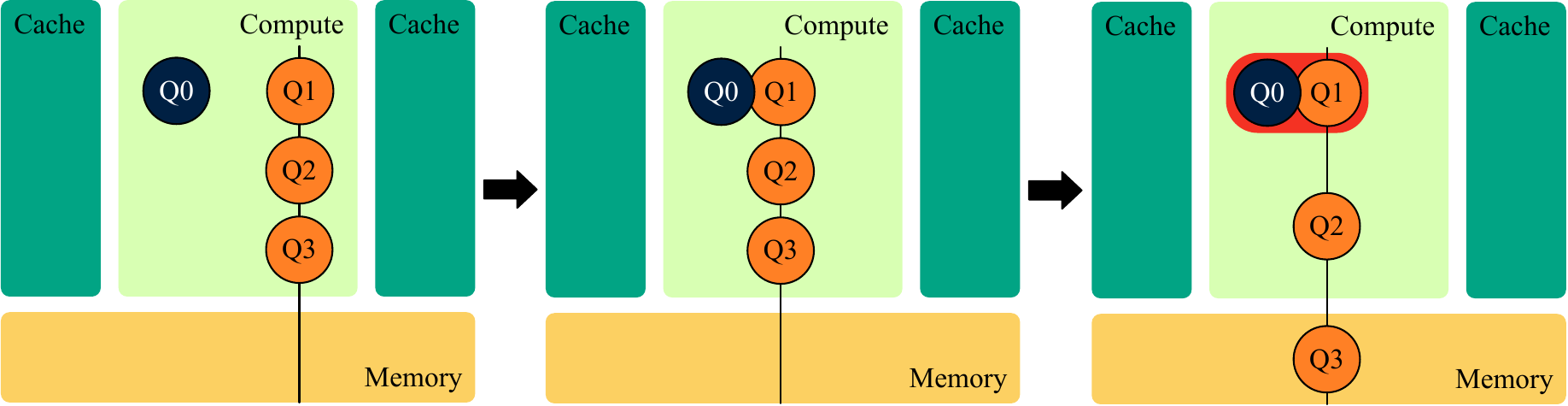}
    \vspace{1mm}
    \hrule
    \vspace{-3.5mm}
    \caption{To run a CZ gate on Q0-Q1, the AOD column containing Q1 is brought closer to Q0, and the other qubits in the same AOD column (Q2 and Q3) are spread apart.}
    \vspace{-1mm}
    \label{fig:expand}
\end{figure}

\subsection{Dual Quantum Cache}

\rev{As discussed previously, crosstalk is a significant concern with quantum computers: qubits unintentionally interacting with each other is a common source of error. Thus, we wish to give mobile qubits that are in the AOD columns the ability to move out of compute and into cache or memory zones as much as possible when the column is not in use. This is to avoid crosstalk of the atoms in the column with other atoms either in the SLM or in the AOD. In order to have more space for AOD columns to avoid crosstalk (e.g., space to move out of compute), \sol{} implements a \textit{dual quantum cache}, where the cache is distributed into two different zones that flank compute on either side (see Fig.~\ref{fig:two_caches}). By having two cache zones, AOD columns have more opportunity to avoid compute when not involved in CZ gates, and thus reduce crosstalk effects. While the addition of this cache does decrease the size of the compute zone, \sol{} does not require a large compute zone as its careful placement of the SLM qubits into the compute zone eliminates the need for atoms to be moved far away within compute to avoid crosstalk.} There is a second reason to have a dual quantum cache, which is that AOD columns can be compiled in a way that ensures no column is given preferential treatment when executing CZ gates, as discussed next. 

\subsection{Quantum Circuit Execution}
\label{sec:exec}

We now discuss how \sol{} compiles the execution routine of a quantum circuit. Recall that after initialization, atoms are in one of two groups; either they are stored in mobile AOD columns, which themselves are in the right cache, or they are stored in static SLM traps in compute.

First, \sol{} checks the left-most AOD column's atoms (in cache) against the SLM atoms, starting with the top-most atom in the AOD column and working down. If and when \sol{} finds an atom in this AOD column that can execute a CZ gate with one of the atoms in compute, \sol{} moves the column such that the two atoms that should entangle are within the interaction radius of each other -- see Q0 in Layer 1 in Fig.~\ref{fig:execution}, which shows \sol{}'s full execution schedule for the example circuit in Fig.~\ref{fig:circuit}(a). Note that when searching for a CZ to execute, we must respect the ordering of the original circuit; although two qubits may have a CZ later in the circuit, this CZ cannot be executed until all previous operations involving the two qubits are completed. Also note that the one-qubit U3 gates (e.g., in Layer 0) can be executed in any zone. Therefore, \sol{} executes those gates regardless of the location of the qubits. \rev{The remainder of the column's atoms get spread out such that they are far enough away from each other to prevent crosstalk.} Note that \sol{} ensures that sufficient space exists for this AOD spreading movement to occur without atoms running out of space in the grid or getting close to other uninvolved atoms in compute. This can be observed in Fig.~\ref{fig:expand} for a sample case.

If none of the atoms in the AOD column need to execute a CZ gate with an atom in compute, then there exists the possibility that one of the atoms in the column needs to execute a CZ with an atom also in the AOD. Recall that we want to guarantee that all CZ gates occur between one qubit in the AOD and one in the SLM. Thus, \sol{} checks the AOD column for any qubits that need to execute a CZ gate with a qubit also in the AOD trap. If any such atoms exist, \sol{} needs to perform a SWAP to move the atom in question into the SLM. \sol{} then attempts to find an atom in the SLM that similarly needs to execute a CZ gate with an atom that is also in the SLM (which also needs to SWAP to execute its CZ). Once such an atom in compute is found, the AOD column is moved such that the two atoms that need to SWAP are within the interaction radius of each other, and like the previous procedure for processing a CZ gate, the remainder of the atoms in the AOD column are spread out to avoid crosstalk (see Layer 5a-j in Fig.~\ref{fig:execution}). If none of the atoms in the AOD column can execute a CZ or a SWAP, then the column must be moved back into the cache to avoid crosstalk and to minimize the amount of movement blocking the column might cause for other columns that still need to compile. This is where having a second cache zone on the left is an asset: \sol{} simply moves the column into this opposite cache zone to avoid having it either block other columns or cause crosstalk with other qubits (see Q0 in Layer 7b-h in Fig.~\ref{fig:execution}).

A challenge that arises is if an AOD column cannot execute any CZ gates or SWAP operations but also cannot move to the opposite-end cache zone because a previously processed AOD column is blocking access due to the relative ordering constraint. Instead of moving to the opposite cache zone, \sol{} solves this challenge by moving the AOD column as close as possible along the X-axis to the previously compiled column and then moving down to the memory zone (see Layer 8b in Fig~\ref{fig:execution}). This also ensures that crosstalk does not occur and that the column does not interfere with the movement of other yet-to-be-processed AOD columns.  With this, the processing of the left-most AOD column is complete, and \sol{} moves on to repeat the entire aforementioned procedure on the second-to-left-most AOD column (which is still in the right cache zone) and continues to repeat until either there are no columns left to process or there is no space for the columns to move into compute. There might not be space if one AOD column entangles with atoms on the right-most side of compute, meaning no other columns can access compute without violating the relative ordering constraint. After all columns are processed, the layer is considered fully compiled, and the Rydberg laser is illuminated to execute all CZ gates simultaneously.

The second layer of CZ gates begins execution by moving all qubits into the opposite cache zone (left) and performing the same steps as above but iterating on AOD columns and atoms in compute from right to left. \sol{} does this to ensure that the left-most AOD columns' qubits do not get arbitrary execution priority over the qubits on the right. Consider if \sol{} moved all AOD columns back to the right cache zone and iterated left-to-right in every layer. The left-most column would always have the freedom to execute CZ gates with any atom in compute; the second-leftmost column would then only be able to perform CZ gates on atoms to the right of the first column (due to the relative ordering constraint) and because of the decreasing accessible compute space each consecutive column would have, the right-most columns would hardly ever get to execute CZ gates until all the qubits in the leftmost columns have completed all of their CZ gates. Switching the initial cache zone and the order in which AOD columns are processed in each layer ensures each AOD column gets similar access to compute per layer. Each layer of CZs thus has qubits starting in the opposite cache zone as the previous layer of CZs. This right-left-right toggle can be seen in Layers 0, 3a, 5a, 5i, 7a, 7i, 8a in Fig.~\ref{fig:execution}.

Between each layer of CZ gates, a layer of U3 gates is executed on all qubits that need to perform U3 rotations. Since U3 rotations can be executed without restrictions via local Raman pulses, qubits anywhere in the system can be accessed, whether in compute, cache, or memory. This can be observed in Layers 0, 2, 4, and 6 in Fig.~\ref{fig:execution}. This leads to an interwoven compilation structure, where alternating U3 and CZ layers are executed until no gates are left in the circuit. 

\subsection{Preemptive SWAPs}

\rev{One challenge that arises is that SWAPs are each composed of three CZ gates and six U3 gates, which means that if \sol{} fully executed SWAPs alongside single gates in a layer, there would be a delay where the qubits executing single gates would need to wait for the SWAPs in the same layer to execute all three gates. This would considerably increase the amount of time it takes to execute each layer in the circuit.}

\rev{To preempt this, \sol{} introduces the idea of breaking each SWAP into its components (three CZ gates and six U3 gates) such that it executes one gate per layer, partially completing the SWAP each time the qubit in the AOD can access the qubit in compute it is swapping with until all gates have completed. For example, if a Q0 in an AOD column has to execute a SWAP with Q1 in compute, Q0 performs the first gate (U3) in the first layer, the second gate (CZ) in the next layer, and so on. By splitting each SWAP into its component gates when possible, \sol{} ensures the runtime of each layer is not impacted by the SWAP's longer execution time.}

\begin{figure}[t]
    \centering
    \includegraphics[scale=0.45]{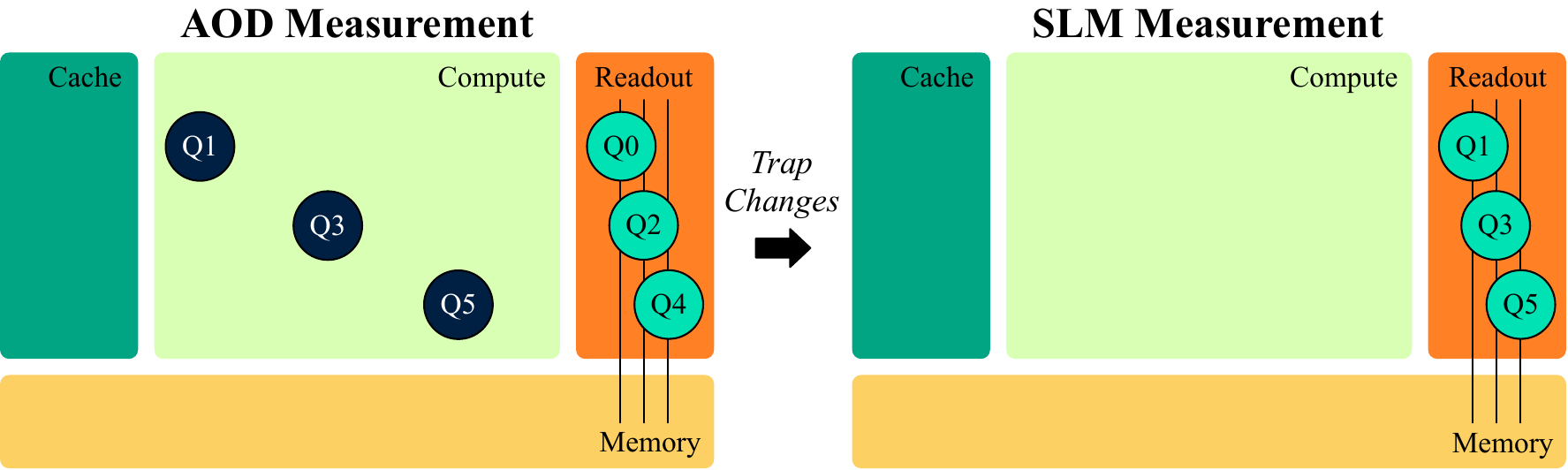}
    \vspace{1mm}
    \hrule
    \vspace{-3.5mm}
    \caption{Last step for the example circuit in Fig.~\ref{fig:circuit}(a). Perform trap changes if needed, move all qubits to the readout zone (same as the left cache), and measure the qubits.}
    \vspace{-1mm}
    \label{fig:measure}
\end{figure}

\subsection{Qubit Measurement}

As shown in Fig.~\ref{fig:measure}, upon completion of circuit execution, all qubits must be moved into the readout zone (same as the right cache) to be measured. \sol{} first moves all qubits in the AOD to the readout zone, and measurement of those qubits is performed via the fluorescence-sensing camera. Then, the AOD columns move to align with all atoms in the compute's SLM, and they pick up all atoms from the SLM in one parallel trap change. Recall that this trap change can happen in parallel since we initially deposited the SLM atoms using the AOD the same way during initialization. These atoms are then moved into the readout zone and measured. This concludes the algorithm. 

\subsection{\rev{Design Trade-offs}}

\rev{While \sol{}'s design offers significant advantages, it is important to acknowledge the trade-offs inherent in its approach.} \rev{\sol{} employs several greedy heuristics throughout its design, most notably the MaxCut algorithm for initial qubit placement (Algorithm~\ref{algo:maxcut}). These heuristics contribute to \sol{}'s ability to achieve fast compilation times, allowing \sol{} to compile large quantum circuits quickly. Yet, this MaxCut algorithm will likely not produce globally optimal results in terms of selecting qubits to be placed in the SLM vs. AOD. While a more exhaustive optimization strategy could potentially improve performance, it would come at the cost of significantly increased compilation time since solving the MaxCut problem optimally is NP-hard, which would make the compilation process intractable for large quantum circuits. The current design of \sol{} thus prioritizes compilation speed and scalability.}

\rev{\sol{}'s design favors SWAP operations over trap changes for moving qubits between the AOD and SLM during circuit execution. This choice is based on the current hardware characteristics, where trap changes are significantly more time-consuming than SWAP operations. However, this design decision also presents a trade-off. By minimizing the use of trap changes, \sol{} reduces the overall circuit runtime and potential decoherence effects that could occur during long trap change operations. However, if future hardware advancements significantly reduce the time required for trap changes, \sol{}'s current approach might not be optimal. It's worth noting that the design of \sol{} allows for adaptation to potential hardware improvements. If trap change times were to decrease substantially, a variant of \sol{} that utilizes trap changes instead of SWAP operations could be implemented with minimal modifications to the overall framework. This flexibility ensures that \sol{} can evolve alongside advancements in hardware.}

\rev{In conclusion, while these trade-offs exist, \sol{}'s design strikes a balance between compilation speed, circuit execution time, and adaptability to hardware changes.}

\subsection{Putting it all Together}
\label{sec:complex}

Overall, \sol{} has a \textit{$O(G[Q_{AOD}Q_{SLM} + QG])$} time complexity, where $\textit{G}$ is the number of gates, $\textit{Q}$ is the total number of qubits, $Q_{AOD}$ is the number of qubits in the AOD, and $Q_{SLM}$ is the number of atoms in the SLM. The $Q_{AOD}*Q_{SLM}$ term is derived from when the compiler checks each atom in the AOD against each atom in the SLM to see where entanglement can occur in each layer. The $QG$ term is from the end of the compilation process for each layer, where \sol{} finds the next gate for each qubit that executed a gate in the previous layer.

Finally, the outer $G$ term is from the possible number of layers of compilation that can be performed. Since \sol{} will never perform more than one SWAP in order to execute a single CZ gate, the number of SWAP gates is necessarily bounded by the number of CZ gates. Thus, the compilation layers that will be executed the most are a factor in the number of CZ gates. It should be noted that Algorithm~\ref{algo:maxcut} does not contribute to the aforementioned upper bound since the greedy MaxCut algorithm iterates through all the gates and has a complexity of $O(G)$.

\rev{It is critical to note that \sol{}'s design is centered around minimizing crosstalk during parallel execution of CZ gates. This is achieved through careful atom placement and management. In the compute zone, atoms in SLM traps are always spaced sufficiently far apart to prevent crosstalk effects when Rydberg excitation occurs. This spacing ensures that SLM atoms do not experience unintended interactions during excitation. When AOD atoms are moved into the compute zone to execute CZ gates with SLM atoms, any uninvolved AOD atoms are strategically positioned either in unoccupied areas of the compute zone or completely outside it. This positioning prevents these uninvolved atoms from causing or experiencing crosstalk.}

\rev{Overall, \sol{}'s lightweight compilation overhead enables it to compile quantum algorithms with over one hundred qubits within one second and with minimal crosstalk (discussed in Sec.~\ref{sec:evaluation}).}

\vspace{1mm}

We now conclude our design discussion. Next, we describe \sol{}'s implementation and evaluation methodology.

\section{Implementation and Methodology}
\label{sec:methodology}

\begin{table}[t]
    \centering
    \caption{\rev{Architecture parameters used for our evaluation.}}
    \vspace{-3.5mm}
    \scalebox{1.0}{
    \begin{tabular}{cl|cl}
         \textbf{Parameter} & \textbf{Value} & \textbf{Parameter} & \textbf{Value} \\
         \hline
         \hline
         Number of Qubits & 51-129 & Atom Loss Rate & 0.7\%~\cite{bluvstein2022quantum} \\
         Trap Change Time & 125$\mu$s~\cite{bluvstein2024logical} & U3 Gate Error & 0.0127\%~\cite{PhysRevA.105.032618} \\
         AOD Movement Speed & 55$\mu$m$/\mu$s~\cite{bluvstein2022quantum} & U3 Gate Time & 2$\mu$s~\cite{Wintersperger2023} \\
         T1 Coherence Time & 4.0s~\cite{bluvstein2022quantum} & CZ Gate Error & 0.48\%~\cite{highfidbluv} \\
         T2 Coherence Time & 1.49s~\cite{bluvstein2022quantum} & CZ Gate Time & 0.8$\mu$s~\cite{bluvstein2022quantum} \\
         SWAP Gate Error & 1.51\%~\cite{highfidbluv} &  Readout Error & 5\%~\cite{Wintersperger2023} \\
         \hline
         \hline
    \end{tabular}}
    \vspace{-1mm}
    \label{tab:params}
\end{table}

\begin{table}[t]
    \centering
    \caption{\rev{Algorithms and benchmarks used for evaluation.}}
    \vspace{-3.5mm}
    \scalebox{1.0}{
    \begin{tabular}{ccl}
         \textbf{Algorithm} & \textbf{Num. Qubits} & \textbf{Description} \\
         \hline
         \hline
         BV & 70 & Bernstein-Vazirani algorithm~\cite{bernstein1993quantum} \\
         CAT & 65 & Coherent superposition of two coherent states~\cite{monroe1996schrodinger} \\
         DNN & 51 & Quantum neural network~\cite{stein2022quclassi}  \\
         GHZ & 78 & Greenberger-Horne-Zeilinger state~\cite{zheng1997preparation} \\
         ISN & 98 & Ising model simulation~\cite{labuhn2016tunable} \\
         ISL & 420 & Large Ising model simulation~\cite{labuhn2016tunable} \\
         KNN & 129 & Quantum $k$ nearest neighbors algorithm~\cite{dang2018image} \\
         QFT & 63 & Quantum Fourier transform~\cite{namias1980fractional} \\
         QGAN & 111 & Quantum generative adversarial network~\cite{dallaire2018quantum} \\
         QV & 1000 & IBM's quantum volume benchmark~\cite{li2023qasmbench} \\
         SWP & 115 & Swap test to measure quantum state distance~\cite{foulds2021controlled} \\
         TFIM & 128 & Transverse-field ising model~\cite{pfeuty1970one} \\
         WST & 76 & W-State preparation and assessment~\cite{fleischhauer2002quantum} \\
         \hline
         \hline
    \end{tabular}}
    \vspace{-1mm}
    \label{tab:algos}
\end{table}

\noindent\textbf{Architectural Simulator Design.} For the size and constraints of our system, we use many of the same figures that were obtained in ~\cite{bluvstein2024logical}. We assume that our atom grid is 370$\mu m$ in width and 190$\mu m$ in height and that the grid can contain up to 280 atoms. The zones used in \sol{} included the two cache zones each of size 80$\mu m$$\times$130$\mu m$ (one of which serves as a readout zone as well), a memory zone of size 190$\mu m$$\times$50$\mu m$, and a compute zone of size 190$\mu m$$\times$130$\mu m$. \rev{For the two highest-qubit-count circuits tested (QV and ISL), each dimension size is doubled to allow for enough room to fit the qubits needed to run the algorithms. Thus, these zones include two cache zones each of size 160$\mu m$$\times$260$\mu m$, a memory zone of size 380$\mu m$$\times$100$\mu m$, and a compute zone of size 380$\mu m$$\times$260$\mu m$.} Crosstalk in our simulation can be avoided when atoms are a distance of 10$\mu m$ apart. To ensure sufficient spacing is available to avoid crosstalk, we limit at most 4 atoms to be in any one AOD column at a time. A Rydberg interaction between atoms occurs when they are less than 2$\mu m$ apart from each other in compute. Atoms that are outside of compute are stored at a distance of 2$\mu m$ apart, whether in the AOD or SLM. For more figures on error and execution times, see Table~\ref{tab:params}. In the aforementioned table, T1 and T2 decoherence times represent the characteristic times for energy relaxation and phase coherence of quantum states, respectively. Together, they represent the ``lifespan'' of a qubit. T1 indicates how long a system can remain in an excited state before returning to the ground state, while T2 measures how long quantum superpositions can be maintained before environmental interactions cause dephasing. T1 and T2 are incorporated into the error rates for our evaluation.

\vspace{2mm}

\noindent\textbf{Compiler Implementation.} We used Python 3.11.5 to implement \sol{}, and Qiskit 0.45.0, IBM's quantum computing framework~\cite{aleksandrowiczqiskit}, is used to decompose quantum circuits into U3 and CZ gates to ensure a Rydberg-atom-compatible representation. Each quantum algorithm is read from a QASM 2.0 file into Qiskit, which contains the circuit represented in the QASM quantum assembly description language. For each input QASM file, we ran the file through the Qiskit transpiler and used the transpiler's output circuit to obtain results for \sol{} and all comparable methods.

In terms of our compiler's code design, we organize all of the major data structures into arrays that have the same number of elements as the number of qubits in the circuit. For instance, we maintain a frontier list that points to the next executable gate for each qubit. Whenever a layer is executed, \sol{} just needs to update the list for whichever qubits executed a gate to point to the next relevant gate. This makes it low overhead to keep track of the gate dependencies so all gates execute in order. Similarly, \sol{} uses hash tables to map qubits to atoms and vice versa. This allows for lightweight updates to qubit-atom mappings when performing SWAPs since it only requires switching the mappings. 

We now describe how the preemptive SWAPs are handled. Recall that, unlike the U3 and CZ gates, SWAPs in \sol{} are not single-layer operations but rather are phased across layers. \rev{Until all three CZ gates and six U3 gates in a SWAP are run, the two qubits involved in a SWAP cannot perform any other gates. Therefore, \sol{} maintains a lookup table of swap statuses, one status per qubit, which counts the number of gates executed so far for the SWAP. Once the counter entry for each qubit is incremented, those qubits get ``locked'' such that they cannot perform any other gates until all SWAP-related gates are executed, at which point the SWAP executes, and the entry is reset.} In a given layer, when a qubit is swapping with a qubit in the opposite type of trap (SLM vs. AOD) to execute a CZ gate, it is vital that the second qubit involved in the CZ gate does not also begin to perform a SWAP, else both qubits might end up entering an infinite loop of SWAPs to get onto opposite types of traps as each other. The locks in the lookup table help ensure that this does not happen. Regarding how SWAP atoms are selected, \sol{} always prefers to SWAP an atom with another that either has no gates left to execute or has to perform a SWAP itself. This minimizes the number of SWAPs \sol{} needs to perform.

We compiled \sol{} on an Apple Mac running the Sonoma 14.2.1 operating system on an Apple M1 chip with 8 cores and 8 GB RAM. We chose this setup to demonstrate the speed of our compiler on a non-server-class computer.

\vspace{2mm}

\noindent\textbf{Quantum Algorithms.} \rev{See Table~\ref{tab:algos} for the complete list of quantum algorithms used for evaluation. The list consists of a diverse set of algorithms of sizes 51-1000 qubits. To stress-test \sol{}, we also evaluate it for the 1000-qubit QV circuit with 5.5 million gates.} The algorithms are taken from the commonly-used QASMBench~\cite{li2023qasmbench} and ArQtiC~\cite{bassman2021arqtic} quantum benchmarking suites. 

\vspace{2mm}

\noindent\textbf{Evaluated Metrics.} \textbf{\textit{(1) Circuit Runtime}} is the end-to-end execution time of a quantum algorithm, including gate and movement times. \textbf{\textit{(2) Estimated Success Probability (ESP)}} is a widely-used metric to calculate the success probability of running a quantum circuit in a noisy quantum environment~\cite{patel2023graphine,li2022optimal,brandhofer2023optimal,tannu2019ensemble,xie2021mitigating}. It is calculated by taking a product of the success rates (which are the complement of the error rates) of individual gates/operations. Also incorporated into the product are the errors based on the T1 and T2 coherence times of the qubits. \textbf{\textit{(3) Number of SWAPs}} represents the overhead of additional gates that have to be added to establish qubit interactions. \textbf{\textit{(4) Number of Trap Changes}} counts the serial trap changes required during the execution of a quantum algorithm. \textbf{\textit{(5) Compilation Time}} is the time it takes to compile the full movement/gate schedule for a given quantum computing algorithm.

\vspace{2mm}

\noindent\textbf{Comparative Techniques.} No direct comparison exists for \sol{} as \sol{} targets the novel zonal addressing architecture~\cite{bluvstein2024logical}. Note that \sol{} cannot be compared against the compilation technique by Bluvstein et al.~\cite{bluvstein2024logical}, as that technique cannot compile general quantum algorithms (any of the algorithms shown in Table~\ref{tab:algos}). Nonetheless, we compare \sol{} against three other techniques. \textbf{\textit{(1) DegreeSplit}} decides the grouping of an atom (SLM or AOD) based on the interaction strength of the qubit -- its degree in the weighted graph. Qubits with the highest degrees are placed in the AOD so that they can travel to different SLM qubits for CZ gates. The rest are placed in the SLM. The remaining procedures are the same as \sol{}. This is the primary technique we compare \sol{} against.

\begin{figure}[t]
    \centering
    \includegraphics[scale=0.544]{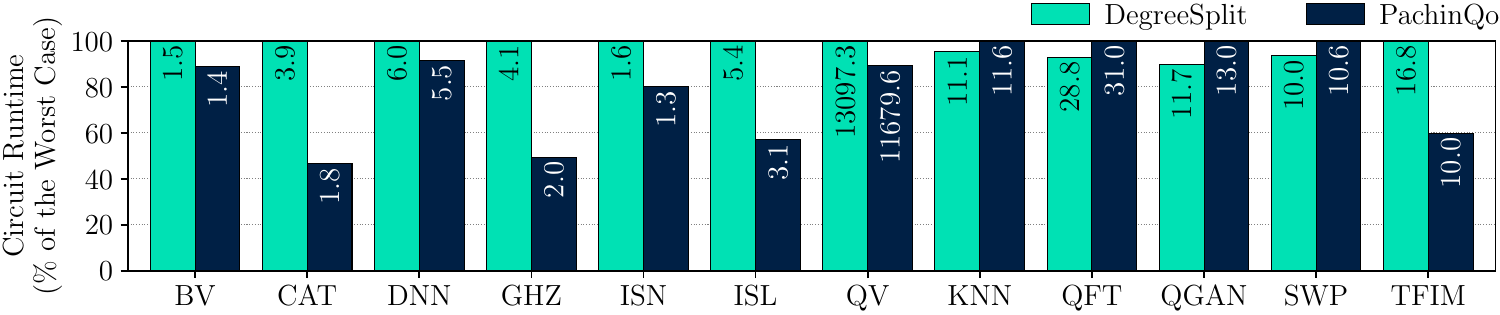}
    \vspace{1mm}
    \hrule
    \vspace{-3.5mm}
    \caption{\rev{\sol{} reduces the average circuit runtime by 20\% over DegreeSplit and outperforms it in most cases. The numbers inside the bars indicate the raw runtimes in $ms$. While the circuit runtimes are short, any decrease improves ESP significantly due to qubit state decoherence.}}
    \vspace{-3mm}
    \label{fig:main_runtime}
\end{figure}

Next, \textbf{\textit{(2) OneCache}} is a variant of \sol{}'s architecture and has only one cache on the right instead of the two caches on either side of compute. The AOD columns from the cache all approach the SLM atoms, perform the CZ gates, and return to their original positions in the cache. This is crudely similar to the proposal by Wang et al.~\cite{wang2023q} (atoms approaching and returning to the same position), but this work is geared toward global addressing architectures. OneCache uses \sol{}'s compiler optimizations. \textbf{\textit{(3) TrapChange}} uses \sol{}'s dual cache architecture (it has the same architecture as \sol{}), but it uses trap changes during circuit execution instead of always using SWAPs. We compare with this technique to show the efficacy of \sol{}'s compiler.
\section{Evaluation and Analysis}
\label{sec:evaluation}

\begin{figure}[t]
    \centering
    \includegraphics[scale=0.544]{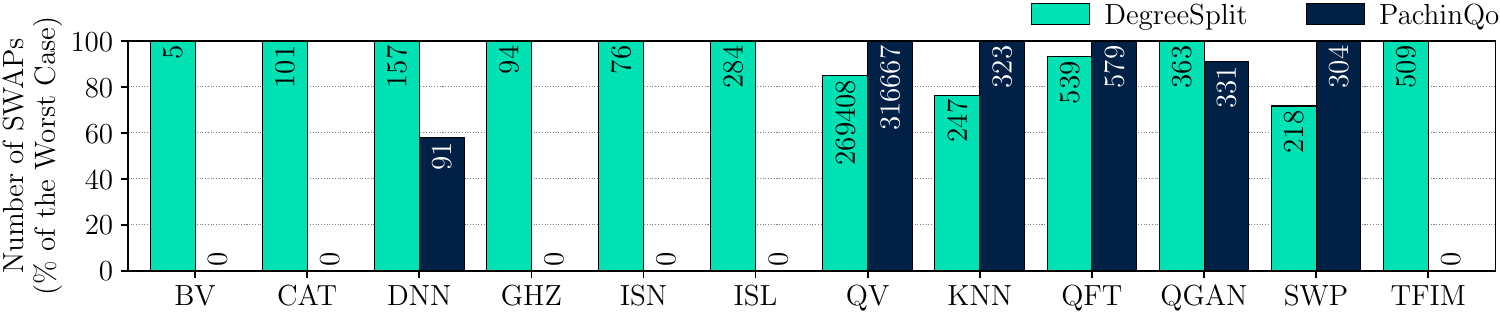}
    \vspace{1mm}
    \hrule
    \vspace{-3.5mm}
    \caption{\rev{\sol{} reduces the avg. num. of SWAPs by 50\% over DegreeSplit and outperforms in most cases.}}
    \vspace{-1mm}
    \label{fig:main_swaps}
\end{figure}

\begin{figure}[t]
    \centering
    \includegraphics[scale=0.544]{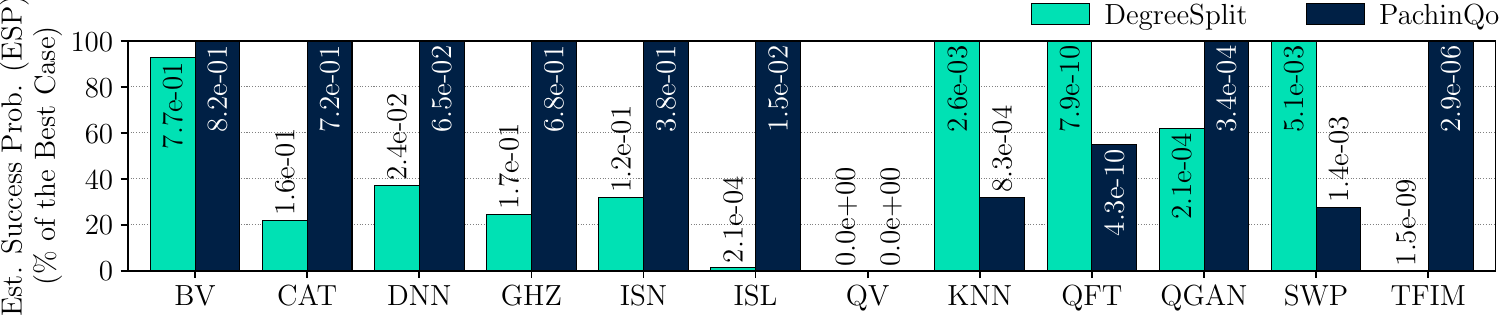}
    \vspace{1mm}
    \hrule
    \vspace{-3.5mm}
    \caption{\rev{DegreeSplit has a 31\% lower estimated probability of success on average as compared to \sol{}. The numbers corresponding to the bars show the raw ESP values.}}
    \vspace{-1mm}
    \label{fig:main_esp}
\end{figure}

\subsection{Primary Results and Analysis}

\noindent \rev{\textbf{Compared to DegreeSplit, \sol{} has 20\% lower circuit runtime, 50\% fewer SWAP gates, and 45\% higher estimated success probability on average.}} As Fig.~\ref{fig:main_runtime} shows, \sol{} has a lower runtime in most of the benchmark algorithms tested when compared to DegreeSplit, having nearly half the runtime in some cases. \rev{This is because, while DegreeSplit needs SWAP gates to execute every algorithm, in roughly half of the benchmarks tested, \sol{} can finish circuit execution without the need for a single SWAP (Fig.~\ref{fig:main_swaps}).} Because of the lower SWAP count and lower runtime, \sol{} has higher ESP than DegreeSplit in most algorithms tested, as shown in Fig.~\ref{fig:main_esp}. Recall that the difference between \sol{} and DegreeSplit is in how the atoms are initially distributed between the SLM traps in compute and the AOD traps. Unlike \sol{}, DegreeSplit places the qubits with the most CZ gates in the AOD. The reason that this strategy does not work well in practice compared to \sol{}'s greedy selection is that most quantum circuits have clusters of regional qubit interactions. For example, in the 128-qubit TFIM circuit, each qubit only interacts with two other qubits, making it effective to partition TFIM's qubits into the SLM and AOD groups with \sol{}.

A similar pattern exists in the circuit for the Ising algorithms, where each qubit executes many CZs but with only one or two other qubits. The implication is that in many algorithms, qubits do not execute CZ gates uniformly with all other qubits in the circuit but rather have a high degree of interactivity with a small subset of qubits. Thus, it is more important that the qubits that interact frequently are placed in separate devices (e.g., AOD vs. SLM in compute). DegreeSplit is completely agnostic about how qubits interact with each other since it only measures the number of CZ gates each qubit is involved in. This leads to it dividing the qubits into the AOD and compute SLM poorly, and thus means it needs more SWAP gates to execute circuits. More SWAP gates mean more CZ gates, which in turn lead to higher operational error and higher runtime. Conversely, since \sol{} has fewer SWAP gates, it has a better performance than DegreeSplit.

Given the advantages that \sol{} has in managing quantum resources, next, we explore how its dual-zone cache architecture plays an important role in reducing overall runtime and enhancing the performance metrics when compared to the single-zone approach of OneCache.

\begin{figure}[t]
    \centering
    \includegraphics[scale=0.544]{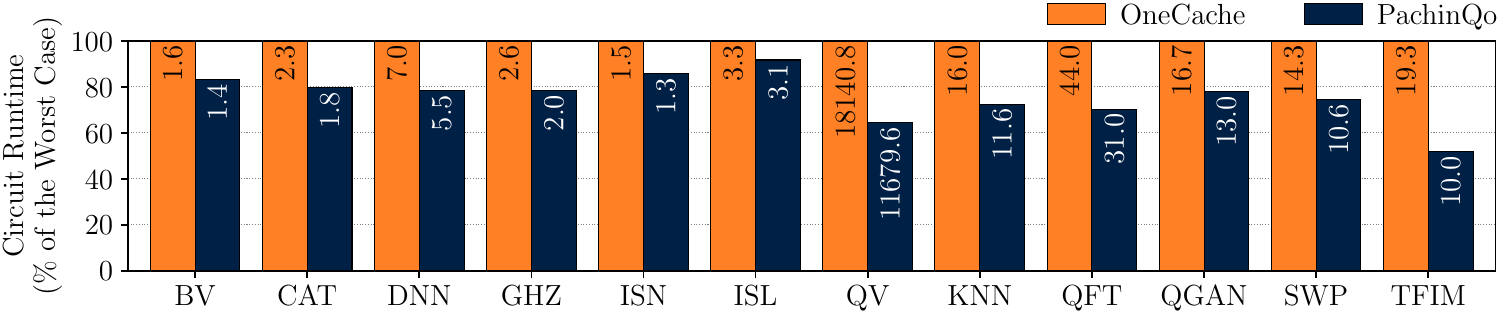}
    \vspace{1mm}
    \hrule
    \vspace{-3.5mm}
    \caption{\rev{\sol{} improves the avg. circuit runtime by 24\% over OneCache and outperforms it in all cases.}}
    \vspace{-1mm}
    \label{fig:one_runtime}
\end{figure}

\subsection{Architecture Comparison: OneCache}

\noindent\textbf{\rev{On average, \sol{}'s dual cache architecture yields a 24\% lower runtime when compared to the single-zone cache alternative, OneCache.}} There are two reasons for why OneCache performs worse than \sol{} (Fig.~\ref{fig:one_runtime}). First, having two caches results in less movement. \rev{Across all algorithms, \sol{} has 32.3\% less total movement than OneCache on average (not depicted in a figure).} Here, total movement is calculated by summing up the distances of all the atom movements that take place during the execution of a quantum circuit.

Second, recall that having a second cache ensures that one side of the AOD (e.g., columns on the left) does not get preferential treatment when it comes to gate compilation. Having such arbitrary preference for certain atoms to compile leads to a longer critical path in the execution of the circuit's gates. For example, the leftmost AOD column's atoms might continuously execute CZ gates with qubits in compute, which is the case with TFIM. While these left-column qubits are engaged, though, other qubits are required to wait, stalling their progress on respective CZ gates. This stalling effect can lead to an uneven distribution of computational tasks, thereby lengthening the runtime of the circuit in some cases.

\rev{While \sol{}'s two-zone cache outperforms OneCache in circuit runtime, the two techniques perform similarly in terms of the ESP: both have an average ESP within 6\% of each other. We, therefore, do not plot this comparisons. These observations emphasize the benefits of \sol{}'s dual-cache architecture. The design choice, as discussed in Sec.~\ref{sec:design}, provides more space for AOD columns to avoid crosstalk when not involved in CZ gates and ensures balanced access to the compute zone for all columns. This approach also addresses the stalling effect observed in OneCache, where left-side AOD columns might monopolize compute access. By preventing such uneven task distribution, \sol{} achieves more balanced parallelism in gate execution. While the dual-cache and single-cache designs perform similarly in terms of SWAP count and ESP, the 24\% reduction in average circuit runtime make the dual-cache design a vital component of \sol{}'s architecture.} Although \sol{} demonstrates a higher performance in circuit runtime due to its efficient cache management, we must also consider other factors, such as how qubits switch between AOD and SLM devices, which are addressed in our comparison with TrapChange, focusing on the use of SWAP gates versus trap changes (discussed next).

\begin{figure}[t]
    \centering
    \includegraphics[scale=0.544]{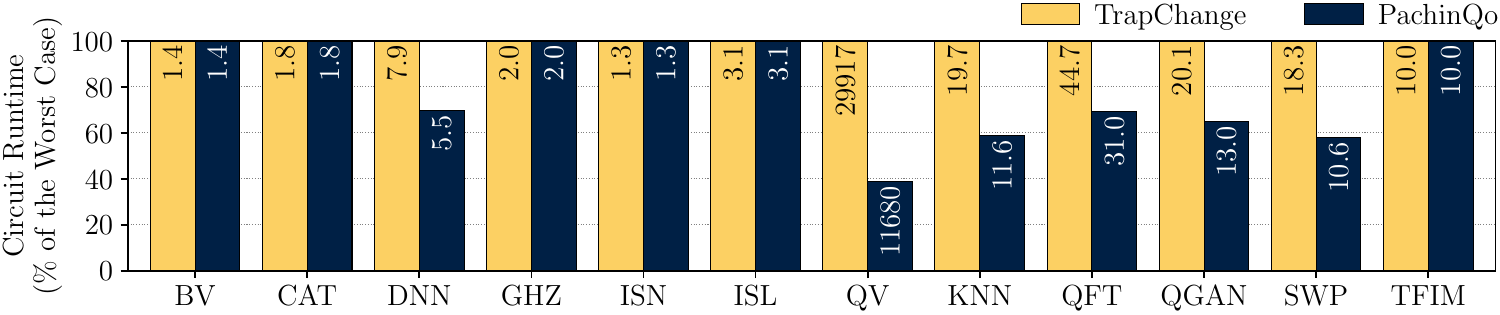}
    \vspace{1mm}
    \hrule
    \vspace{-3.5mm}
    \caption{\rev{\sol{} improves the avg. circuit runtime by 19\% over TrapChange as it reduces serial trap changes.}}
    \vspace{-1mm}
    \label{fig:trap_runtime}
\end{figure}

\begin{figure}[t]
    \centering
    \includegraphics[scale=0.544]{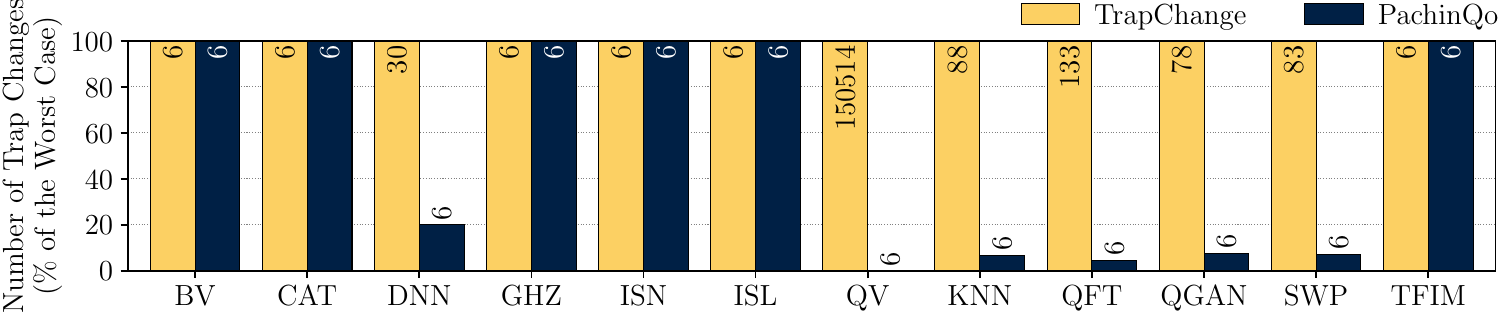}
    \vspace{1mm}
    \hrule
    \vspace{-3.5mm}
    \caption{\rev{While some algorithms with few trap changes may not benefit from \sol{}'s SWAP operations, \sol{}'s use of SWAP operations helps reduce trap changes by 43\% on average, which is especially beneficial for algorithms with many trap changes.}}
    \vspace{-1mm}
    \label{fig:trap_changes}
\end{figure}

\subsection{Compiler Comparison: TrapChange}

\noindent\textbf{\rev{Compared to TrapChange, \sol{} has 19\% lower runtime and 43\% fewer trap changes across all algorithms}}. Refer to Fig.~\ref{fig:trap_runtime} and Fig.~\ref{fig:trap_changes} for the results. While it is unsurprising that \sol{} has significantly fewer trap changes than TrapChange, the results show the runtime impact of having SWAP gates instead of trap change operations to switch qubits from one atom trap to another. \rev{Recall from Table~\ref{tab:params} that trap change operations are highly time-consuming, with it taking about 52$\times$ longer to perform a trap change than a SWAP gate (three CZ gates and six U3 gates).} This is the cause for TrapChange's significantly longer circuit runtimes than \sol{} for many of the algorithms. It should be noted that \sol{} only has six trap changes for all algorithms (similarly, DegreeSplit and OneCache also only have six trap changes for all algorithms). These occur as a result of the initialization of the atom grid and of the final qubit measurement at the end of circuit execution. During initialization, \sol{} needs to perform three trap changes to move atoms into compute's SLM and the AOD: two trap changes to get atoms from memory into AOD and then from AOD into SLM, and a third to move the remaining atoms in memory into the AOD. Similarly, after circuit completion, measurement requires three trap changes. These six trap changes are required regardless of the algorithm. Thus, the number of trap changes remains consistent with \sol{}. 

Having discussed the efficacy of \sol{} in using SWAP gates versus trap changes, let us now examine how \sol{} performs for different SLM qubit arrangements.

\subsection{\sol{}'s Performance with Respect to SLM Qubit Grids}

\begin{figure}[t]
    \centering
    \subfloat[Large Square Grid]{\includegraphics[width=0.24\linewidth]{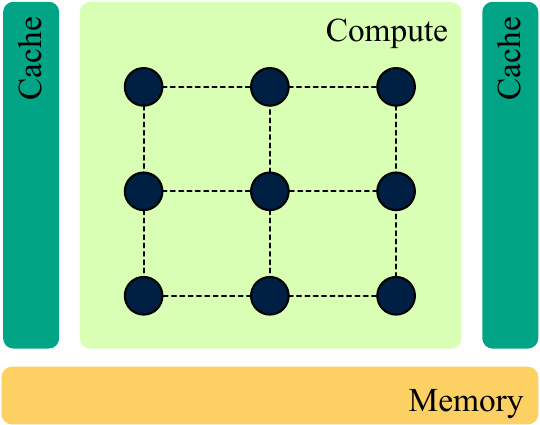}}
    \hfill
    \subfloat[Small Square Grid]{\includegraphics[width=0.24\linewidth]{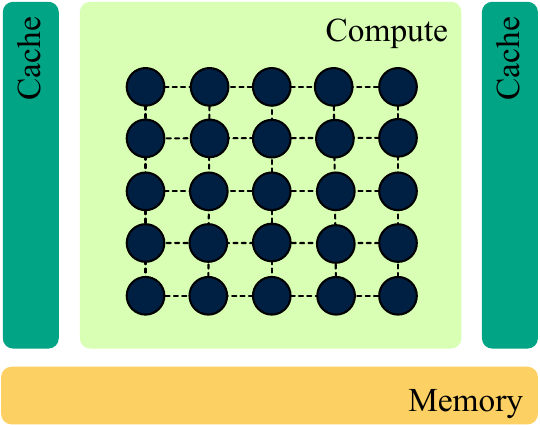}}
    \hfill
    \subfloat[Traingle Grid]{\includegraphics[width=0.24\linewidth]{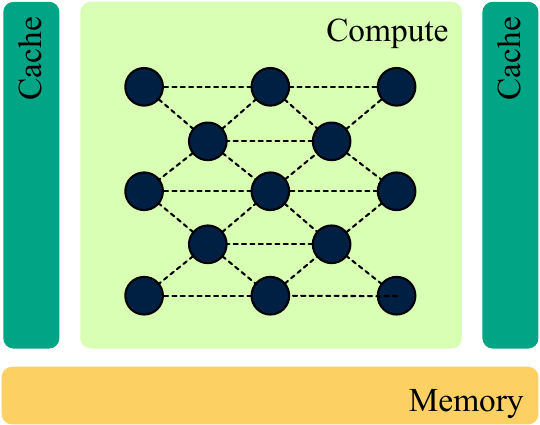}}
    \hfill
    \subfloat[Star Grid]{\includegraphics[width=0.24\linewidth]{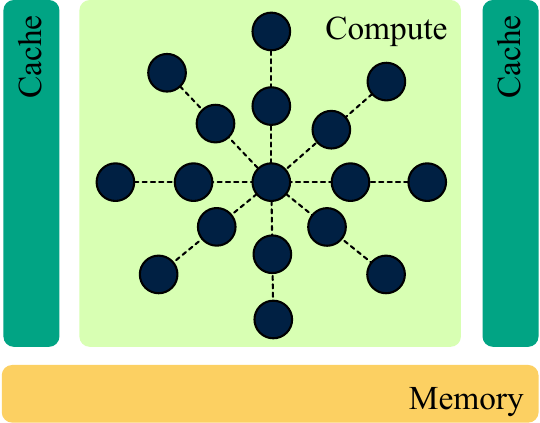}}
    \vspace{1mm}
    \hrule
    \vspace{-3.5mm}
    \caption{We explore the four types of grids depicted for SLM qubit arrangement. (a) The large square grid enables all neighboring qubits to be equidistant from each other and ensures that they are far enough to be outside of each other's interaction zones. This also enables AOD atoms to line up in columns for parallel operations. We use this grid as the default grid for \sol{}. (b) The small square grid has all the benefits of the large square grid but can fit more atoms in the SLM, enabling more parallelization in cases when the AOD atoms in a column line up perfectly for operations with the SLM atoms in a column. However, it can cause more serialization if atoms in different AOD columns need to interact with the SLM atoms in the same column. (c) The triangle grid can potentially offer a balance between the large square grid and the small square grid by having an in-between sparsity. (d) Lastly, the star grid provides the ability to have some SLM qubits sparsely arranged (those near the edges) while having others densely arranged (those near the center), which could be beneficial for algorithms with interactions of varying sparsity.}
    \vspace{-1mm}
    \label{fig:grids}
\end{figure}

\begin{figure}[t]
    \centering
    \includegraphics[scale=0.517]{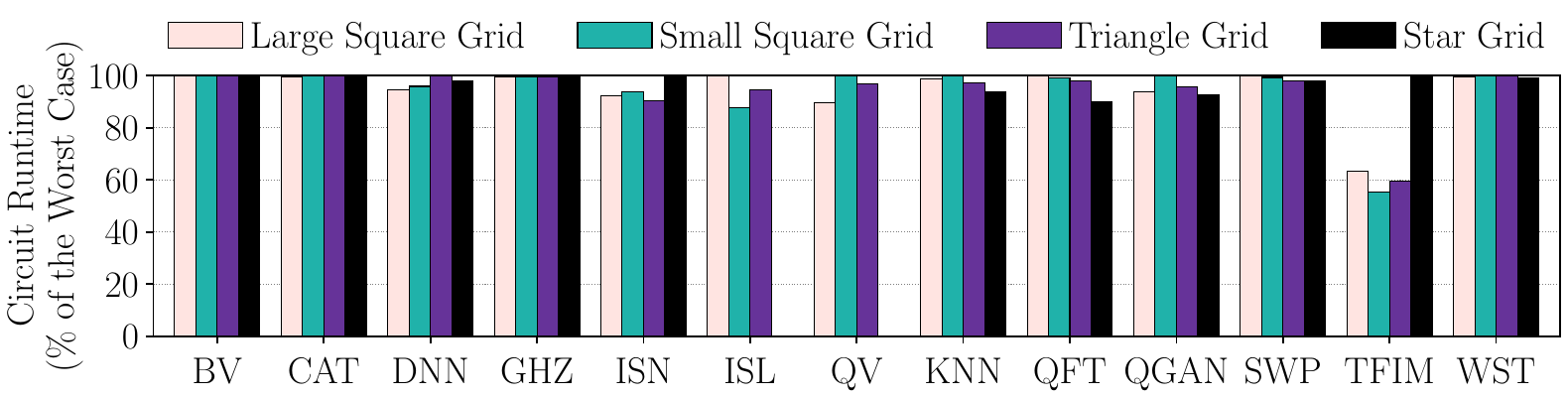}
    \vspace{1mm}
    \hrule
    \vspace{-3.5mm}
    \caption{\rev{\sol{} achieves similar circuit runtimes on average regardless of the chosen SLM arrangement grid. Nonetheless, some algorithms, such as TFIM, can benefit from being matched to compatible SLM arrangement grids (e.g., any grid other than the star grid). Note that it was not possible to run ISL/QV on the Star Grid configuration due to lengthy compilation times for that grid.}}
    \vspace{-1mm}
    \label{fig:runtime_gap}
\end{figure}

\textbf{On average, \sol{}'s performance remains largely unimpacted by the chosen SLM atom arrangement, requiring no tuning effort to optimize the compiler for different grids.} We evaluate the performance of \sol{} with respect to several atom arrangement grids for the SLM qubits to demonstrate the impact of the SLM qubit arrangement. The grids we examined and the rationale behind selecting them are presented in Fig.~\ref{fig:grids}. We use the large square grid as the default configuration for \sol{}. We evaluate two metrics to assess the performance impact of the different grids: (1) circuit runtime and (2) ESP.

Fig.~\ref{fig:runtime_gap} presents the results for circuit runtime. The results demonstrate that most algorithms do not get notably impacted by the SLM atom arrangement grid. Thus, the average circuit runtime remains similar regardless of the SLM atom grid chosen. This is due to the fact that \sol{}'s compiler design inherently adapts to the needs of the chosen architecture configuration so that it can perform well for different grids for most algorithms. One exception is the TFIM algorithm, which performs considerably better with all grids other than the star grid. This is due to the ``staircase'' entangling structure of the TFIM circuit, which is more conducive to grid structures with patterns of fixed sparsity, which is the case with all grids except for the star grid.

\begin{figure}[t]
    \centering
    \includegraphics[scale=0.515]{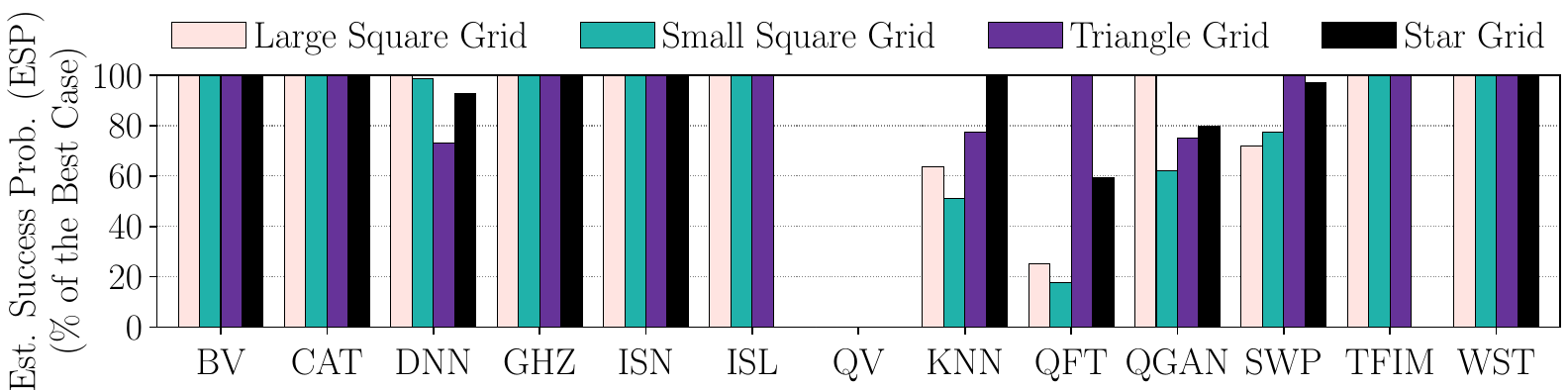}
    \vspace{1mm}
    \hrule
    \vspace{-3.5mm}
    \caption{\rev{\sol{} achieves mostly similar ESP results on average for different SLM arrangement grids. However, the large square grid and the triangle grid perform better for some algorithms and make a substantial difference for algorithms like DNN, QFT, QGAN, and SWP. Note that all bars for QV are at zero due to the ESP being negligible on account of 5.5 million gates.}}
    \vspace{-1mm}
    \label{fig:esp_gaps}
\end{figure}

\rev{On the other hand, Fig.~\ref{fig:esp_gaps} shows the results for the ESP metric. While all the grids still perform similarly on average, the large square grid and the triangle grid perform slightly better. This is because for algorithms like SWP and QFT, the triangle grid performs the best, while for algorithms like DNN and QGAN, the large square grid performs the best. Matching the grid to their circuit structure can make a marked difference for these algorithms.} Nonetheless, because the large square grid performs well in general, we choose it as the default for \sol{}, but \sol{} can be run with any other grid without requiring any compilation tuning effort, which highlights the general applicability and customizability of the \sol{} framework.

Having explored the impact of different SLM grid types \sol{}'s performance, we lastly discuss how \sol{} achieves fast compilation times.

\begin{table}[t]
    \centering
    \caption{\rev{Compilation times of algorithms with \sol{}. Note that we report the median here as the average would be skewed by QV's 5.5 million gates.}}
    \vspace{-3.5mm}
    \scalebox{1.0}{
    \begin{tabular}{cll|cll}
         \textbf{Algoritm} & \textbf{Num. Gates} & \textbf{Comp. Time} & \textbf{Algoritm} & \textbf{Num. Gates} & \textbf{Comp. Time} \\
         \hline
         \hline
         BV & 178 & 12.6$ms$ & QFT & 6498 & 709.1$ms$ \\
         CAT & 193 & 26.8$ms$ & QGAN & 1695 & 344.7$ms$ \\
         DNN & 765 & 79.7$ms$ & QV & 5.5M & 4.7 $h$ \\
         GHZ & 232 & 36.2$ms$ & SWP & 1085 & 296.4$ms$ \\
         ISN & 534 & 56.0$ms$ & TFIM & 6526 & 804.3$ms$ \\
         ISL & 3356 & 1.6 $s$ & WST & 451 & 55.1$ms$ \\
         KNN & 1218 & 367.9$ms$ & \textbf{Median} & 1085 & 296.4 $ms$ \\
         \hline
         \hline
    \end{tabular}}
    \vspace{-1mm}
    \label{tab:comp_times}
\end{table}

\subsection{\sol{}'s Compilation Overhead}

\rev{\textbf{\sol{} achieves milli-second compilation times for most evaluated algorithms.}} Table~\ref{tab:comp_times} provides \sol{}'s compilation times for all the algorithms. \rev{As the table shows, \sol{} has sub-second compilation times across most algorithms, even the TFIM algorithm with 129 qubits. The only exceptions are the ISL algorithm (420 qubits, 3.4K gates), which compiled in 1.64 $s$, and the QV algorithm (1K qubits, 5.5M gates), which took 4.7 $h$.} \rev{Excluding the QV algorithm, the average compilation time with \sol{} is 366 $ms$.} As the table indicates, the compilation times are correlated with the number of gates in the quantum algorithm, which is consistent with the time complexity described in Sec.~\ref{sec:complex}. This is a result of the deliberate choices made in the design of \sol{} to make compilation algorithm decisions based on lightweight but well-performing heuristics such as greedy MaxCut, dual cache right-left-right movements, and preemptive SWAPs. This was an important consideration for \sol{} as these compilers need to be able to scale to future quantum computing systems with hundreds and thousands of qubits, which will also run algorithms with hundreds and thousands of qubits. \sol{}'s compiler is already designed with this future scalability in mind.
\section{Related Work}
\label{sec:related_work}

\textbf{Rydberg Atom Quantum Architecture Design.} Much of the architecture development effort has focused on individual and global addressing systems~\cite{wurtz2023aquila,bluvstein2022quantum,PhysRevA.105.032618,Wintersperger2023,highfidbluv,shi2020single}. As an instance, Shi et al.~\cite{shi2020single} design an architecture for individual addressing, while the QuEra Aquila quantum computer~\cite{wurtz2023aquila} proposes an architecture for analog quantum computing with global addressing. To strike a balance between the individual and global addressing architectures, Bluvstein et al.~\cite{bluvstein2024logical} pave the way toward zonal addressing Rydberg atom quantum computers with separate zones for compute, memory, and readout. They demonstrate this architecture on a Rydberg atom system but only for one type of program. One contribution of \sol{} is to build on this architecture work by developing the dual-cache architecture to accommodate general quantum algorithms.

In previous works, the assumed setup for Rydberg atom computers involved a 2-dimensional AOD, which can store rows and columns of mobile qubits but with the constraint that qubits in a single row or column must move together~\cite{wurtz2023aquila,bluvstein2022quantum, grahamgatemodel}. This constraint has been one of the major limitations when designing parallelized neutral atom compilation techniques since the implication is that moving one AOD qubit would coincidentally move an entire array of qubits. In contrast, with \sol{}, we use a paradigm similar to a recent work by Adams et al.~\cite{adamscustomaod} that shows an architecture where the AOD is composed of columns of qubits where each qubit can move independently of all other qubits in one dimension. We utilize this recently developed capability in the design of \sol{} to motivate the ability for qubits in each AOD column to move independently of one another (with ordering constraints maintained).

\vspace{2mm}

\noindent\textbf{Rydberg Atom Quantum Compiler Design.} Several works have proposed compilers for Rydberg atom quantum computers~\cite{Baker2021-wv,litteken2022reducing,patel2022geyser,patel2023graphine,tan2022qubit,tan2023compiling,wang2023q,bluvstein2024logical}. These works either target individual addressing architectures or global addressing architectures. For example, Patel et al.~\cite{patel2023graphine} develop customizable atom placement without atom movement for algorithm-specific needs for individual addressing quantum computers. On the other hand, Tan et al.~\cite{tan2022qubit} propose a solver-based approach for scheduling the movements of atoms on global addressing quantum computers. 

There are only two compilers designed for zonal addressing architectures. The first was proposed by Bluvstein et al.~\cite{bluvstein2024logical} upon their introduction of the zonal architecture. However, this compiler is only designed for surface codes, which have all parallel qubit gate operations, and thus is not generally applicable. Similarly, recently, a work by Stade et al.~\cite{stade2024abstractmodelefficientrouting} also designed a compiler for zonal architectures but also focused on compiling for error-corrected qubits, which have parallel operations and, therefore, is not suitable for general algorithms run on near-future quantum computers. The works also do not propose a hardware-software co-design framework.

\vspace{1mm}

In contrast to the above prior art, \sol{} co-designs architecture and compilation for zonal addressing Rydberg atom computers to execute any given quantum algorithm.
\section{Conclusion}
\label{sec:conclusion}

In this paper, we detailed \sol{}, a first-of-its-kind framework for the co-design of architecture and compilation for zonal addressing Rydberg atom quantum computers for general quantum algorithms. \sol{} addresses novel and non-trivial challenges in these Rydberg atom systems by developing a dual quantum cache, a greedy MaxCut initialization routine, and an execution schedule that leverages careful movement and preemptive/parallelizable SWAPs. \rev{\sol{} improves the estimated probability of success by 45\% on average.} We believe that \sol{} will lead the way for future hardware-software co-design of Rydberg atom computers.

\section*{Acknowledgement}

We would like to thank the anonymous reviewers of the ACM SIGMETRICS 2025 Conference and our shepherd, Professor Mahmut Kandemir, for their valuable and insightful feedback. This work was supported by Rice University and the Department of Computer Science at Rice University. This work was also supported by the Ken Kennedy Institute and Rice Quantum Initiative, which is part of the Smalley-Curl Institute.

\balance

\bibliographystyle{ACM-Reference-Format}
\bibliography{main}

\end{document}